\documentclass[12pt,tightenlines,eqsecnum,floats,aps,amsmath,amssymb,nofootinbib,prd,showpacs]{revtex4}

\usepackage{setspace}
\usepackage{amsmath,amssymb,amsfonts,amsthm}
\usepackage{graphicx}
\usepackage{enumerate}

\newcommand{\dos}{{\scriptstyle{(2)}}}
\newcommand{\tres}{{\scriptstyle{(3)}}}
\newcommand{\cuatro}{{\scriptstyle{(4)}}}

\begin{document}
\title{Hamiltonian Dynamics of Linearly Polarized Gowdy
 Models Coupled to Massless Scalar Fields}

\author{J. Fernando \surname{Barbero G.}}
\email[]{fbarbero@iem.cfmac.csic.es} \affiliation{Instituto de
Estructura de la Materia, CSIC, Serrano 123, 28006 Madrid, Spain}
\author{Daniel  \surname{G\'omez Vergel}}
\email[]{dgvergel@iem.cfmac.csic.es} \affiliation{Instituto de
Estructura de la Materia, CSIC, Serrano 123, 28006 Madrid, Spain}
\author{Eduardo J. \surname{S. Villase\~nor}}
\email[]{ejsanche@math.uc3m.es} \affiliation{Grupo de
Modelizaci\'on y Simulaci\'on Num\'erica, Universidad Carlos III
de Madrid, Avda. de la Universidad 30, 28911 Legan\'es, Spain}
\affiliation{Instituto de Estructura de la Materia, CSIC, Serrano
123, 28006 Madrid, Spain}

\date{July 23, 2007}

\begin{abstract}
The purpose of this paper is to analyze in detail the Hamiltonian
formulation for the compact Gowdy models coupled to massless
scalar fields as a necessary first step towards their quantization. We will
pay special attention to the coupling of matter and those features that arise for the
$\mathbb{S}^1\times\mathbb{S}^2$ and $\mathbb{S}^3$ topologies that are
not present in the well studied $\mathbb{T}^3$ case --in
particular the polar constraints that come from the regularity
conditions on the metric. As a byproduct of our analysis we will
get an alternative understanding, within the Hamiltonian
framework, of the appearance of initial and final
singularities for these models.
\end{abstract}


\maketitle

\section{Introduction}{\label{Intro}}

Symmetry reductions are a way to gain useful insights for difficult
problems in classical and quantum general relativity. In this
respect the two Killing vector reductions provided by the so called
Gowdy models are specially attractive because they have a
cosmological interpretation and share some interesting features with
similar reductions such as the Einstein-Rosen waves --in particular
their solvability-- both at the classical and quantum regimes. Most
of the work on these models, after the initial papers by Gowdy
\cite{Gowdy:1971jh,Gowdy:1973mu}, has profusely analyzed those
corresponding to the 3-torus spatial topology and, in fact, this is
by far the preferred choice to discuss quantization issues
\cite{Misner,Berger:1975kn,Berger:1973,Cortez,
Corichi:2006xi,Corichi:2006zv,Mena:1997,Torre:2002xt,Torre:2007zj,
Romano:1996ep,Corichi:2002vy, Pierri:2000ri,BarberoG.:2006zw}. The
other possible closed (compact and without boundary) topologies, the
\textit{three-handle} $\mathbb{S}^1\times\mathbb{S}^2$, the
three-sphere $\mathbb{S}^3$, and the lens spaces $L(p,q)$, are
interesting in their own right. From the physical point of view
their most salient feature is the fact that they describe
cosmological models with both initial and final singularities. For
this reason they will become useful test beds for issues related to
quantization in cyclic universes.

The first step towards quantization is the Hamiltonian formulation
of the model at hand. This is specially so for constrained systems
where the identification of the relevant constraints is a
necessary first step either to attempt a phase space reduction, a
gauge fixing, or a quantization \textit{\`a la Dirac} where the
physical Hilbert space is identified as the kernel of suitable
self-adjoint operators representing the constraints.

Our goal in this paper is to perform a detailed Hamiltonian analysis
of the compact Gowdy models coupled to massless scalar fields,
extending in several ways previous results on this subject
\cite{BarberoG.:2005ge,BarberoG.:2006gd}. Adding matter fields to
the system is a way to enrich these models and get closer to
physically realistic situations. Here we will work in the spirit of
previous treatments for other two Killing vector reductions
\cite{Ashtekar:1996bb,Beetle:1998iu}, paying close attention to the
constraint analysis, gauge fixing, and deparameterization. To our
knowledge the Hamiltonian analysis, for the vacuum
$\mathbb{S}^1\times\mathbb{S}^2$ and $\mathbb{S}^3$ Gowdy models,
has only been addressed in a partial way in \cite{Hanquin} where the
authors give Hamiltonians for these systems. However they do not
provide the detailed phase space description (constraints, gauge
fixing, and so on) necessary to understand relevant geometrical
issues. Also their reduced phase space treatment does not allow to
follow other roads to quantization such as Dirac's approach or the
viewpoint pioneered by Varadarajan in \cite{Varadarajan:2006am}.
Among several issues we want to find out how the topology of the
spatial slices affects the definition of the constraints, and how
the coupling of massless scalar fields is realized in the different
topologies. Along the way we also want to understand in a detailed
way, and within the Hamiltonian setting, the mechanisms leading to
the appearance of final singularities. Our results can be
immediately particularized to the vacuum situation.

The paper is structured as follows. After this introduction we will
review in section \ref{generalframework} the main points concerning
the Geroch reduction for polarized Gowdy models coupled to massless
scalar matter fields. In particular we will show that this reduction, and a
subsequent conformal transformation, allows us to interpret these
models as 2+1 gravity coupled to a set of massless scalar fields
with axial symmetry. Some details related to this reduction vary
depending on the topologies (in particular those related to the
quotient spatial manifold) and will be commented separately for each case.

Section \ref{Torus} will be devoted to discuss the Gowdy
$\mathbb{T}^3$ models coupled to massless scalar fields extending
the previous treatments for the vacuum case
\cite{Pierri:2000ri,Mena:1997} (and similar models such as
Einstein-Rosen waves \cite{BarberoG.:2005ge,BarberoG.:2006gd}).
Along the way we will clarify some issues related to
deparameterization and the appearance of singularities in the $3+1$
dimensional metrics. This section will be the basis of the treatment
that we will follow to study the other possible topologies.

Section \ref{handle} will be devoted to the Hamiltonian
formulation of Gowdy models in $\mathbb{S}^1\times\mathbb{S}^2$
coupled to massless scalars. Here we will have to pay special
attention to the identification of the regularity conditions that
the basic fields describing the model must satisfy as a
consequence of the regularity conditions on the metric.
As we will see the constraints that are relevant here
are different from the ones present for the 3-torus due
to the presence of a symmetry axis in the spatial manifold. In particular
we will get what we will refer to as ``polar constraints''
involving the values of the basic fields at the poles of the two
dimensional sphere that appears as the quotient space after
performing a Geroch reduction. As we will show they are first class and play a
relevant role to guarantee the differentiability of the other
constraints. Another issue that will be discussed is how the deparameterization
achieved by a partial gauge fixing works for this model and how one can arrive
at a reduced phase space description. We will see that, as it also
happens in the $\mathbb{T}^3$ case, the dynamics of the system is
described by a time dependent Hamiltonian though the time
dependence now is different and reflects the appearance of initial
and final singularities. In fact, as a result of our analysis, we
will get a geometric understanding of this fact in terms of the
geometry of the constraint hypersurface in phase space.

After this we will perform a similar analysis in section \ref{S3}
for the $\mathbb{S}^3$ topology. Here the main difference stems
from the fact that we will be forced to perform the Geroch
reduction needed to describe the model in $2+1$ dimensions by
using a Killing field whose norm vanishes on a circle
$\mathbb{S}^1$. This will introduce some modifications in our
description and will change the analysis of the relevant
regularity conditions for the metric. Nevertheless we will find
out that the final description is quite similar to the one
corresponding to the three handle discussed above.

The detailed quantization of the $\mathbb{S}^1\times\mathbb{S}^2$
and $\mathbb{S}^3$ Gowdy models will be carried out elsewhere. A
fact that will play a relevant role there is the possibility of
describing the compact Gowdy models in the different topologies as
field theories in certain conformally stationary curved
backgrounds. As this point of view is also useful to understand
some of the issues discussed in the paper from a different
perspective we will show in  section \ref{back} how this can be done.

We end the paper in section \ref{conclusions} with a
discussion of the main results and suggestions for future work on this
subject.

\section{General features of compact Gowdy models:
 Geroch reduction and $2+1$ dimensional
 formulation}{\label{generalframework}}

Let us consider a smooth, effective, and proper action of the
biparametric Lie Group $G^\dos:=U(1)\times
U(1)=\{(g_1,g_2)=(e^{ix_1},e^{ix_2})\,|
\,x_{1},x_{2}\in\mathbb{R}(\mathrm{mod}\,2\pi)\}$ on a compact,
connected, and oriented 3-manifold ${^{\scriptstyle{\tres}}}\Sigma$.
It can be shown \cite{MOSTERT,Chrusciel:1990zx} that this action is
unique up to automorphisms of $G^\dos$ and diffeomorphisms of
${^{\scriptstyle{\tres}}}\Sigma$. The spatial manifold
${^{\scriptstyle{\tres}}}\Sigma$ is then restricted to have the
topology of a three-torus $\mathbb{T}^3$, a three-handle
$\mathbb{S}^1\times \mathbb{S}^2$, the three-sphere $\mathbb{S}^3$,
or the lens spaces $L(p,q)$ (that can be studied by imposing discrete
symmetries on the $\mathbb{S}^3$ case).

Let us take a four manifold ${^{\cuatro}}\mathcal{M}$ diffeomorphic
to $\mathbb{R}\times{^{\tres}}\Sigma$ and such that
$({^{\cuatro}}\mathcal{M}, {^{\cuatro}}g_{ab})$ is a globally
hyperbolic spacetime endowed with a Lorentzian
metric\footnote{Throughout the paper we will use the Penrose
abstract index convention  with tangent space indices belonging to
the beginning of the Latin alphabet \cite{Penrose}.  Lorentzian
spacetime metrics will have signature $(-+++)$ and the conventions
for the curvature tensors are those of Wald \cite{Wald}.}
${^{\cuatro}}g_{ab}$. Let us further require that $G^\dos$ acts by
isometries on the spatial slices of ${^{\cuatro}}\mathcal{M}$. In
this paper we will focus on the so called \emph{linearly polarized
case}, hence, the isometry group will be generated by a pair of
mutually orthogonal, commuting, spacelike, and globally defined
hypersurface-orthogonal Killing vector fields
$(\xi^{a},\sigma^{a})$.

Let us consider now the Einstein-Klein-Gordon
equations
\begin{equation}\label{EKG}
{^{\cuatro}}R_{ab}=8\pi
G_{N}(\mathrm{d}\phi)_a(\mathrm{d}\phi)_b\,,\quad
{^{\cuatro}}g^{ab}\,{^{\cuatro}}\nabla_{a}{^{\cuatro}}\nabla_{b}\phi=0\,
\end{equation}
corresponding to (3+1)-dimensional gravity minimally
coupled to a zero rest mass scalar field $\phi$ symmetric
under the diffeomorphisms generated by the Killing fields
($\mathcal{L}_{\xi}\phi=\mathcal{L}_{\sigma}\phi=0$,
$\mathcal{L}_{\xi}{^{\cuatro}}g_{ab}=\mathcal{L}_{\sigma}{^{\cuatro}}g_{ab}=0$).
Here ${^{\cuatro}}R_{ab}$ and ${^{\cuatro}}\nabla_{a}$ denote the
Ricci tensor and  the metric connection  associated to
${^{\cuatro}}g_{ab}$, respectively. The exterior derivative of the
scalar field $\phi$ is denoted by $(\mathrm{d}\phi)_a$ and  $G_{N}$
is the Newton constant.

In order to get a simplified, lower dimensional description we will
perform a Geroch reduction \cite{Geroch:1970nt} by taking advantage
of the existence of Killing vector fields. The possibility of
finding the necessary non-vanishing Killing field $\xi^a$ will
depend, as we will see later, on the spatial topology that we
consider. In some cases the appropriate Killing vectors vanish on
2-dimensional submanifolds but, nevertheless, we will be able to use
Geroch's procedure even in this situation. The idea is to find a
suitable reduction on the manifold
${^{\tres}}\mathcal{M}={^{\cuatro}}\tilde{\mathcal{M}}/U(1)$,
diffeomorphic to $\mathbb{R}\times{^{\dos}}\Sigma$,  where
${^{\cuatro}}\tilde{\mathcal{M}}$ denotes the set of points in
${^{\cuatro}}\mathcal{M}$ in which $\xi^{a}$ is nonvanishing, and
reintroduce the removed points (the symmetry axis) as a boundary
where the fields must satisfy certain regularity conditions. In the
present situation hypersurface orthogonality will allow us to view
${^{\tres}}\mathcal{M}$ as an embedded submanifold,
everywhere orthogonal to the closed orbits of $\xi^a$, and endowed
with the induced metric
${^{\tres}}g_{ab}:={^{\cuatro}}g_{ab}-\lambda_{\xi}^{-1}\xi_{a}\xi_{b}$,
where
$\lambda_{\xi}:={^{\cuatro}}g_{ab}\xi^{a}\xi^{b}:=\xi_{a}\xi^{a}>0$.

In the linearly polarized case the twist of the Killing fields vanishes and
the field equations can be written as those corresponding to a set
of massless scalar fields coupled to (2+1)-gravity by performing the conformal
transformation $g_{ab}:=\lambda_{\xi}{^{\tres}}g_{ab}$. The
system (\ref{EKG}) is then equivalent to
\begin{equation}
R_{ab}=\frac{1}{2}\sum_i
(\mathrm{d}\phi_i)_a(\mathrm{d}\phi_i)_b\,,\quad
g^{ab}\nabla_a\nabla_b\phi_i=0\,,\quad \mathcal{L}_\sigma
g_{ab}=0\,,\quad \mathcal{L}_\sigma\phi_i=0\,,\label{ecs}
\end{equation}
where $R_{ab}$ and $\nabla_{a}$ denote, respectively, the Ricci
tensor and the Levi-Civita connection associated to $g_{ab}$ (all of
them three dimensional objects), we have defined\footnote{2+1
massless scalar fields will be denoted be the subindex $i=1,2$. The
subindex $i=1$ will label the gravitational scalar that encodes the
local gravitational degrees of freedom in Gowdy models and the
subindex $i=2$ will label the original 3+1 matter scalar. It is
completely straightforward to couple any number $N$ of massless
scalar fields, in practice this can be done by supposing that the
index $i$ runs from 1 to $N$.} $\phi_1:=\log\lambda_{\xi}$,
$\phi_2:=\sqrt{16\pi G_{N}}\phi$, and we must remember that we
have the additional symmetry generated by the remaining Killing
vector field $\sigma^a$. Notice that (\ref{ecs}) are formally
symmetric under the exchange of the gravitational and matter
scalars. However, it is important to realize that for some of the
topologies that we will discuss, these fields may be subject to
different regularity conditions in the gravitational and matter
sectors that effectively break the symmetry among them. The relevant
details for each spatial topology will be given in the corresponding section.

In order to obtain the Hamiltonian formulations for the models that
we are considering here we want to derive the previous equations
from an action principle. To this end we introduce the
(2+1)-dimensional Einstein-Hilbert action corresponding to gravity
coupled to massless scalars
\begin{eqnarray}\label{act}
\displaystyle {^{\tres}}S(g_{ab},\phi_i)&=&\frac{1}{16\pi
G_{3}}\int_{(t_0,t_1)\times^{\dos}\Sigma}\!\!\!^{\tres}\mathrm{e}\,|g|^{1/2}\left(R-
\frac{1}{2}\sum_i g^{ab}(\mathrm{d}\phi_i)_a(\mathrm{d}\phi_i)_b\right)\nonumber\\
& & \displaystyle +\frac{1}{8\pi
G_{3}}\int_{\{t_{0}\}\times{^{\dos}\Sigma\cup\{t_{1}\}\times{^{\dos}}\Sigma}}
\!\!\!^{\dos}\mathrm{e}\,|h|^{1/2}K\,.
\end{eqnarray}
Here $R$ denotes the Ricci scalar associated to $g_{ab}$. $K$ and
$h_{ab}$ are, respectively, the trace of the second fundamental form
$K_{ab}$ (defined by the exterior normal unit vector $n^a$), and the
induced 2-metric on the boundary $\{t_{0}\}\times^
{\tres}\Sigma\cup\{t_{1}\}\times ^{\tres}\Sigma$. Finally $G_{3}$
denotes the Newton constant per unit length in the direction of the
$\xi$-symmetric orbits. We have restricted the integration region to
an interval $[t_0,t_1]$, where $t$ defines a global coordinate on
$\mathbb{R}$. The action is written with the help of a fiducial
(i.e. non dynamical) volume form ${^{\tres}}\mathrm{e}$ compatible
with the canonical volume form ${^{\tres}}\epsilon$ defined by the
metric $g_{ab}$. This is given by\footnote{In any basis where the
nonvanishing components of ${^{\tres}}\epsilon$ have the values
$\pm1$, $|g|^{1/2}$ is equivalent to the square of the absolute
value of the determinant of the matrix of the metric in that basis.}
${^{\tres}}\epsilon=\sqrt{|g|}\,{^{\tres}}\mathrm{e}$. The volume
form ${^{\tres}}\epsilon$ induces a 2-form
${^{\dos}}\epsilon_{ab}={^{\tres}}\epsilon_{abc}n^{c}$ on each slice
$\{t\}\times{^{\dos}}\Sigma$ that agrees with the volume associated
to the 2-metric $h_{ab}$. We have also introduced a fixed volume
2-form ${^{\dos}}\mathrm{e}$ on $\{t\}\times{^{\dos}}\Sigma$ such
that ${^{\dos}}\epsilon=\sqrt{|h|}\,{^{\dos}}\mathrm{e}$, and
verifies
$\sqrt{|g|}\,{^{\tres}}\mathrm{e}_{abc}n^{c}=\sqrt{|h|}\,{^{\dos}}\mathrm{e}_{ab}$.
We require that ${^{\tres}}\mathrm{e}$ and ${^{\dos}}\mathrm{e}$ be
time-independent, i.e. $\mathcal{L}_{t}{^{\tres}}\mathrm{e}=0$,
$\mathcal{L}_{t}{^{\dos}}\mathrm{e}=0$ where $\mathcal{L}_{t}$
denotes Lie derivative along $t^{a}:=(\partial/\partial t)^{a}$. We
also demand them to be invariant under the action of the remaining
Killing vector field. In particular, given the (2+1)-dimensional
splitting of ${^{\tres}}\Sigma$ it is natural to choose
${^{\tres}}\mathrm{e}=\mathrm{d}t\wedge{^{\dos}}\mathrm{e}$, with
$\mathcal{L}_{t}{^{\dos}}\mathrm{e}=\mathcal{L}_{\sigma}{^{\dos}}\mathrm{e}=0$
\cite{Wald}.

For all the different topologies, using the Stokes theorem, we get
\begin{widetext}
\begin{eqnarray}\label{actr}
\!\!\!\!\displaystyle
{^{\tres}}S(g_{ab},\phi_i)\!=\!\frac{1}{16\pi
G_{3}}\int_{t_0}^{t_1}\!\!\!\mathrm{d}t\!\int_{{^{\dos}}\Sigma}\!\!
\!\!\!^{\dos}\mathrm{e}\,|g|^{1/2}\Bigg(\,^{\dos}R+K_{ab}K^{ab}-K^2
- \frac{1}{2}\sum_i g^{ab}
(\mathrm{d}\phi_{i})_a(\mathrm{d}\phi_{i})_b\Bigg)
\end{eqnarray}
\end{widetext}
where we have used the relation
$R={^{\dos}}R+K_{ab}K^{ab}-K^{2}+2\nabla_a(n^{a}K-n^{b}\nabla_{b}n^{a})$
and ${^{\dos}}R$ denotes the Ricci scalar associated
to $h_{ab}$. Our strategy in the different topologies that we will study in the
paper will be to write down an action of this type, adapted to the
peculiarities of the different spatial topologies (in particular
those originating in the different sets of regularity conditions
that we will have to consider) and use it to derive a Hamiltonian
formulation for the system.

\section{$\mathbb{T}^3$ Gowdy models coupled to massless scalars}{\label{Torus}}

The $\mathbb{T}^3$ Gowdy model is, by far, the most studied to date
both at the classical and quantum levels \cite{Gowdy:1971jh,
Gowdy:1973mu,Misner,Berger:1975kn,Berger:1973,Cortez,
Corichi:2006xi,Corichi:2006zv,Mena:1997,Torre:2002xt,
Torre:2007zj,Romano:1996ep,Corichi:2002vy,
Pierri:2000ri,BarberoG.:2006zw,Chrusciel:1990zx}. We will consider
in this section the coupling of some types of matter fields and the
most important aspects of its Hamiltonian treatment paying special
attention to the deparameterization and reduced Hamiltonian
description.

Let us start by considering the orientable 3-manifold
${^{\scriptstyle{\tres}}}\Sigma=\mathbb{T}^3
=\mathbb{S}^1\times\mathbb{S}^1\times\mathbb{S}^1$, whose points we
parameterize in the form
$(z_1,z_2,z_3)=(e^{i\theta},e^{i\sigma},e^{i\xi})$ with
$\theta,\sigma,\xi\in\mathbb{R}(\mathrm{mod}\,2\pi)$. In particular,
we endow ${^{\scriptstyle{\tres}}}\Sigma$ with the standard volume
form $\mathrm{d}\theta\wedge\mathrm{d}\sigma\wedge\mathrm{d}\xi$. We
define the following (left) $G^\dos$-group action
$$
(g_1,g_2)\cdot(z_1,z_2,z_3)=(e^{ix_1},e^{ix_2})\cdot(e^{i\theta},e^{i\sigma},e^{i\xi})
:=(e^{i\theta},e^{i(x_1+\sigma)},e^{i(x_2+\xi)})\,.
$$
We can consider now the group orbits defined by the commuting
subgroups $(g_1,g_2)=(e^{ix},1)$, $(g_1,g_2)=(1,e^{ix})$,
$x\in\mathbb{R}(\mathrm{mod}\,2\pi)$
\begin{eqnarray*}
&&(e^{ix},1)\cdot(e^{i\theta},e^{i\sigma},e^{i\xi})=(e^{i\theta},e^{i(x+\sigma)},e^{i\xi})\,,\nonumber\\
&&(1,e^{ix})\cdot(e^{i\theta},e^{i\sigma},e^{i\xi})=(e^{i\theta},e^{i\sigma},e^{i(x+\xi)})\,,\nonumber
\end{eqnarray*}
and their corresponding tangent vectors at each point of
$\mathbb{T}^3$ obtained by differentiating the previous expressions
with respect to $x$ at $x=0$
\begin{eqnarray*}
&&(0,ie^{i\sigma},0)\,,\nonumber\\
&&(0,0,ie^{i\xi})\,.\nonumber
\end{eqnarray*}
Let us consider the four manifold
${^{\scriptstyle{\cuatro}}}\mathcal{M}\simeq\mathbb{R}\times\mathbb{T}^3$.
We introduce now three smooth vector fields $\theta^{a}$, $\sigma^a$
and $\xi^a$, tangent to the embedded submanifolds
$\{t\}\times\mathbb{T}^3$ (here $t$ is a global coordinate on
$\mathbb{R}$). In order to do this, let us fix $t_0\in\mathbb{R}$ and define on
$\{t_0\}\times\mathbb{T}^3$ the 3-dimensional vector fields\footnote{Notice that even though we use a coordinate notation these are globally defined vector fields on the spatial manifolds $\{t_0\}\times\mathbb{T}^3$.}
$(\partial/\partial\theta)^{a}$, $(\partial/\partial\sigma)^{a}$,
and $(\partial/\partial\xi)^{a}$ given in the description of ${^{\scriptstyle{\tres}}}\Sigma$ at the beginning of this section. We extend them to
${^{\scriptstyle{\cuatro}}}\mathcal{M}$ by Lie dragging along a
smooth vector field $t^a$ defined\footnote{In particular, take
$t^{a}:=(\partial/\partial t)^{a}$.} as the tangent vector to a
smooth congruence of curves transverse to the slices
$\{t\}\times\mathbb{T}^3$.  Notice that given a one parameter family
of diffeomorphisms $f_t$ we have
$f_{t*}[\xi,\sigma]^a=[f_{t*}\xi,f_{t*}\sigma]^a$ so we guarantee
that the extended fields commute everywhere. The 4-tuple
$(t^{a},\theta^{a},\sigma^{a},\xi^{a})$ defines then a
paralelization of ${^{\scriptstyle{\cuatro}}}\mathcal{M}$. Here $\theta^a$ is the vector field obtained by extending $(\partial/\partial\theta)^a$ to the four-dimensional manifold ${^{\scriptstyle{\cuatro}}}\mathcal{M}$; $\sigma^a$ and $\xi^a$ are obtained by the same procedure. Once we have introduced these vector fields on
${^{\scriptstyle{\cuatro}}}\mathcal{M}$ as \textit{background}
objects we restrict ourselves to working with metrics
$^{\cuatro}g_{ab}$ satisfying the following conditions:

\begin{enumerate}

\item The action of the group $G^\dos$ on
${^{\scriptstyle{\cuatro}}}\mathcal{M}$ defined by
$(g_1,g_2)\cdot(t,p)=(t,(g_1,g_2)\cdot p)$, $t\in\mathbb{R}$,
$p\in\mathbb{T}^{3}$, with $(g_1,g_2)\cdot p$ defined above, is an action
by isometries, i.e. $\xi^a$ and $\sigma^a$ are Killing vector fields
($\mathcal{L}_{\xi}^{\cuatro}g_{ab}=0$,
$\mathcal{L}_{\sigma}^{\cuatro}g_{ab}=0$).

\item $t$ is a global time function, i.e.
${^{\scriptstyle{\cuatro}}}g^{ab}(\mathrm{d}t)_b$ is a timelike
vector field. From now on we will consider the manifold
${^{\scriptstyle{\cuatro}}}\mathcal{M}$ to be endowed with a time
orientation such that this vector field is past-directed.
\item $\{t\}\times\mathbb{T}^{3}$ are spacelike hypersurfaces for all
 $t\in\mathbb{R}$. In particular
  $\lambda_{\xi}:={^{\scriptstyle{\cuatro}}}g_{ab}\xi^a\xi^b>0$,
$\lambda_{\sigma}:={^{\scriptstyle{\cuatro}}}g_{ab}\sigma^a\sigma^b>0$.

\item $\xi^a$ and $\sigma^a$ are hypersurface orthogonal (this
defines the so called linearly polarized case). This condition means
that the twist of the two fields vanishes. This will ultimately
allow us to simplify the field equations and describe the system as
a simple theory of scalar fields.
\end{enumerate}
Two simple but important results that can be proved at this point as
a consequence of the first are the following:

\bigskip

\noindent i) If $\xi^a$ and $\sigma^a$ are Killing
vectors and $[\xi,\sigma]^a=0$ then
$\mathcal{L}_\sigma(^{\cuatro}g_{ab}-\xi_a\xi_b/\lambda_{\xi})=0$;

\bigskip

\noindent ii) Furthermore, if we define the vector $X^a$ orthogonal to $\xi^a$
as $X^a:=\sigma^a-\xi^a(\xi^b\sigma_b)/\lambda_{\xi}$ it satisfies
$[\xi,X]^a=0$ and also
$\mathcal{L}_X(^{\cuatro}g_{ab}-\xi_a\xi_b/\lambda_{\xi})=0$. This
means that, without loss of generality, we can work with everywhere
orthogonal and commuting Killing vector fields $\xi^a$ and
$\sigma^a$. In fact, we impose

\bigskip

\indent 5. $(\theta^{a},\sigma^{a},\xi^{a})$ are mutually
${^{\cuatro}}g$-orthogonal vector fields.

\bigskip

After we perform the Geroch reduction with respect to the field
$\xi^{a}$ as described above we end up with a set of equations that
can be obtained from a 2+1 dimensional action of the type
(\ref{actr}) with
${^{\dos}}\Sigma=\mathbb{T}^2=\mathbb{S}^1\times\mathbb{S}^1$. Since
the remaining Killing vector field $\sigma^{a}$ is still
hypersurface orthogonal, and non-vanishing, the corresponding space of orbits
${^{\dos}}\mathcal{M}:={^{\tres}}\mathcal{M}/U(1)\simeq\mathbb{R}\times\mathbb{S}^{1}$
can be identified as an embedded hypersurface in
$^{\tres}\mathcal{M}$ everywhere orthogonal to the (closed) orbits
of $\sigma^{a}$. The induced 2-metric of signature $(-+)$ on
$^{\dos}\mathcal{M}$ can be written
$$s_{ab}=g_{ab}-\tau^{-2}\sigma_a\sigma_b\,,$$
where
$\tau^2:=g_{ab}\sigma^{a}\sigma^{b}=\lambda_{\xi}\lambda_{\sigma}>0$
is the area density of the symmetry $G^\dos$-group orbits. In the
following we will use the notation $\tau=+\sqrt{\tau^2}$. We have now an induced foliation
over $^{\dos}\mathcal{M}$ defined by the global time function $t$
introduced before. Let $n^{a}$ be the $g$-unit and future-directed
($g^{ab}n_{a}(\mathrm{d}t)_b>0$) vector field normal to this
foliation, and let $\hat{\theta}^{a}$ be the $g$-unit spacelike
vector field of closed orbits tangent to the slices of constant $t$,
such that
$$\theta^{a}=e^{\gamma/2}\hat{\theta}^{a}$$
for some extra field $\gamma$. If we choose the congruence of curves with $t^{a}$
tangent to ${^{\dos}}\mathcal{M}$. Then, the congruence is
transverse to the foliation, and we can express
\begin{equation}\label{ta}
t^{a}=e^{\gamma/2}(Nn^{a}+N^{\theta}\hat{\theta}^{a})\,,
\end{equation}
where $N>0$ and $N^{\theta}$ are proportional to
the lapse and shift functions. The factor
$e^{\gamma/2}$ will allow us to obtain a proper gauge algebra and simplify later calculations.
We require that $N$, $N^{\theta}$,
and $\gamma$ are smooth real-valued fields on
${^{\tres}}\mathcal{M}$. As we will see in the following the
symmetry generated by $\sigma^{a}$ will further constraint them, in
particular they will be constant along the orbits defined by the
remaining Killing vector field. The orthonormal basis
$(n^{a},\hat{\theta}^{a},\sigma^{a}/\tau)$ is positively oriented
with respect to the volume 3-form associated to the 3-metric
$g_{ab}$, compatible with
$\mathrm{d}t\wedge\mathrm{d}\theta\wedge\mathrm{d}\sigma$,
satisfying
${^{\tres}}\epsilon_{abc}n^{a}\hat{\theta}^{a}\sigma^{a}/\tau=1$.

The expression of the metric is
\begin{eqnarray}\label{metric}
&&g_{ab}=e^\gamma\bigg((N^{\theta
2}-N^{2})(\mathrm{d}t)_a(\mathrm{d}t)_b+
2N^{\theta}(\mathrm{d}t)_{(a}(\mathrm{d}\theta)_{b)} +
(\mathrm{d}\theta)_a(\mathrm{d}\theta)_b\bigg)+
\tau^{2}(\mathrm{d}\sigma)_a(\mathrm{d}\sigma)_b\,.\quad\quad
\end{eqnarray}
The fact that the vectors $(t^a,\theta^a,\sigma^a)$
commute everywhere will translate into necessary conditions that the
vectors $n^a$ and $\theta^a$ and the scalars $N$, $N^\theta$, and
$\gamma$ must satisfy. These are
\begin{eqnarray}
\mathcal{L}_\sigma N=0\,,\,&\mathcal{L}_\sigma
N^\theta=0\,,\,&\mathcal{L}_\sigma\gamma=0\,,\label{dersigma}
\end{eqnarray}
\begin{eqnarray} &(\mathcal{L}_\sigma n)^a=0\,,\,&
(\mathcal{L}_\sigma \hat{\theta})^a=0\,,\label{comm}
\end{eqnarray}
and
\begin{eqnarray}
\frac{1}{2}(\mathcal{L}_{\theta}\gamma)(Nn^a+N^\theta\hat{\theta}^a)
-e^{-\gamma/2}(\mathcal{L}_te^{\gamma/2})\hat{\theta}^a+N
e^{\gamma/2}[\hat{\theta},n]^a+(\mathcal{L}_{\theta}N)n^a
+(\mathcal{L}_{\theta}N^\theta)\hat{\theta}^a=0\,.\nonumber
\end{eqnarray}
This last equation can be projected in the directions defined by the
basis vectors to give
\begin{eqnarray*}
&&\frac{1}{2}N\mathcal{L}_\theta\gamma+\mathcal{L}_\theta
N+Ne^{\gamma/2}n^an^b\nabla_a\hat{\theta}_b=0\,,\label{proy_1}\\
&&\frac{1}{2}N^\theta\mathcal{L}_\theta\gamma+\mathcal{L}_\theta
N^\theta-\frac{1}{2}\mathcal{L}_t\gamma+
Ne^{\gamma/2}\hat{\theta}^a\hat{\theta}^b\nabla_a n_b=0\,,\label{proy_2}\\
&&\hat{\theta}^a\sigma^b\nabla_an_b=0\,.\label{proy_3}
\end{eqnarray*}
These equations are important because they relate the components of
the extrinsic curvature of some surfaces with derivatives of $N$,
$N^{\theta}$, and $\gamma$. Notice that the scalars $\phi_i$
are also constant on the orbits of $\sigma^a$ (the matter
scalar $\phi_2$ because we have imposed this from the start and the
gravitational scalar $\phi_1$ due to the fact that the two Killings $\xi^a$
and $\sigma^a$ commute: $\mathcal{L}_\sigma\lambda_{\xi}=0$).
Therefore, as we will end up with an essentially two dimensional
model with fields depending only on coordinates $t$ and
$\theta$, we will eventually denote $\mathcal{L}_t$ with a dot and
$\mathcal{L}_\theta$ with a prime. With this convention, we obtain:
\begin{eqnarray}
{^{\dos}}R&=&\tau^{-1}e^{-\gamma}(\tau'\gamma'-2\tau'')\,,\\
K_{ab}K^{ab}-K^2&=&
-\frac{e^{-\gamma}}{N^2\tau}(\dot{\gamma}
-N^{\theta}\gamma'-2N^{\theta\prime})(\dot{\tau}-N^{\theta}\tau')\,,\\
g^{ab}
(\mathrm{d}\phi_i)_a(\mathrm{d}\phi_i)_b&=&-\frac{e^{-\gamma}}{N^{2}}
(\dot{\phi}_{i}^{2}-2N^{\theta}\dot{\phi}_{i}\phi_{i}^{\prime}
+({N^{\theta}}^{2}-N^2)\phi_{i}^{\prime 2})\,,
\end{eqnarray}
and then the action can be written as
\begin{eqnarray*}
&&\frac{1}{16\pi G_3}\int_{t_0}^{t_1}
\mathrm{d}t\int_{\mathbb{T}^2}{^{\dos}}\mathrm{e}|g|^{1/2}e^{-\gamma}\bigg(
\frac{1}{\tau}(\gamma^\prime\tau^\prime-2\tau^{\prime\prime})-\frac{1}{\tau
N^2} (\dot{\gamma}-2N^{\theta\prime}-\gamma^\prime
N^\theta)(\dot{\tau}-N^\theta\tau^\prime)\nonumber\\
&&\hspace{4cm}+\frac{1}{2N^2}\sum_{i}\big[\dot{\phi}^2_i-2N^\theta\dot{\phi}_i\phi_i^\prime+
(N^{\theta 2}-N^2)\phi^{\prime2}_i\big]\bigg)\,. \nonumber
\end{eqnarray*}
This will be the starting point for the Hamiltonian formalism.
Notice that the previous expression is coordinate independent. The
Lagrangian is written as an integral over the torus $\mathbb{T}^2$
of the 2-form obtained by multiplying the fiducial volume form and a
scalar function. All the terms in this scalar are
defined through the use of geometrical objects, in particular the
derivatives are Lie derivatives along the fields introduced above.
This will prove particularly important when dealing with other
spatial topologies. In this case, it is natural to choose as
fiducial 2-form ${^{\dos}}\mathrm{e}$ the one verifying
${^{\dos}}\mathrm{e}_{ab}\theta^{a}\sigma^{b}=Ne^{\gamma}\tau/|g|^{1/2}=1$,
i.e. ${^{\dos}}\mathrm{e}=\mathrm{d}\theta\wedge\mathrm{d}\sigma$.

The Hamiltonian can be easily obtained by performing a Legendre
transform. It has the form
\begin{eqnarray*}
H=C[N]+C_\theta[N^\theta]\,,
\end{eqnarray*}
where\footnote{Here and in the following $\displaystyle\int_{\mathbb{S}^1}
F:=\int_{\mathbb{S}^1} F \,\mathrm{d}\theta$.}
\begin{eqnarray*}
&&C[N]=\int_{\mathbb{S}^{1}} NC:=\int_{\mathbb{S}^1}
N\bigg(\frac{1}{8G_3}(2\tau^{\prime\prime}-\gamma^\prime\tau^\prime)-8G_3p_\gamma
p_\tau+\frac{1}{2}\sum_{i}\Big(8G_3\frac{p_{\phi_{i}}^2}{\tau}+
\frac{\tau}{8G_3}\phi_{i}^{\prime2}\Big)\bigg)\,,\label{const_N}\\
&&C_\theta[N^\theta]=\int_{\mathbb{S}^{1}}
N^{\theta}C_{\theta}:=\int_{\mathbb{S}^1} N^\theta
\bigg(-2p_\gamma^\prime+p_\gamma \gamma^\prime+p_\tau
\tau^\prime+\sum_{i} p_{\phi_{i}}\phi_{i}^\prime
\bigg)\,.\label{const_N_theta}
\end{eqnarray*}
The lapse and the shift act as Lagrange multipliers and
enforce the constraints $C=0$, $C_{\theta}=0$. The canonical
phase space $(\Gamma,\omega)$ is coordinatized by the canonically
conjugate pairs
$(\gamma,p_{\gamma};\tau,p_{\tau};\phi_{i},p_{\phi_i})$
and $\omega$ denotes the standard (weakly)
symplectic form
\begin{equation}\label{ws2f}
\omega=\int_{\mathbb{S}^{1}}\Big(\delta\gamma\wedge\delta
p_{\gamma}+\delta\tau\wedge\delta
p_{\tau}+\sum_{i}\delta\phi_{i}\wedge\delta p_{\phi_{i}}\Big)\,.
\end{equation}
The dynamical variables are restricted to belong to a constraint submanifold $\Gamma_{c}\subset\Gamma$
globally defined by $C=0$, $C_{\theta}=0$. The constraints can be
written in an equivalent way by taking ``linear combinations''
obtained by integrating them against suitable weight functions
$N_{g}$ and $N^{\theta}_{g}$ in such a way that the vanishing of the
weighted form of the constraints for all of them is equivalent to
the vanishing of $C$ and $C_{\theta}$ at every point of
$\mathbb{S}^1$. The gauge transformations generated by the
(weighted) constraints are\footnote{In the rest of this section we
will choose units such that $8G_3=1$.}
\begin{eqnarray*}
&&\{\gamma,C[N_g]\}=-N_gp_\tau\,,\nonumber\\
&&\{\tau,C[N_g]\}=-N_gp_\gamma\,,\nonumber\\
&&\{\phi_{i},C[N_g]\}=N_g\frac{p_{\phi_i}}{\tau}\,,\nonumber\\
&&\{p_\gamma,C[N_g]\}=-(N_g\tau^\prime)^\prime\,,\nonumber\\
&&\{p_\tau,C[N_g]\}=
-(N_g\gamma^\prime)^\prime+\frac{1}{2}N_g\sum_{i}\Big(\frac{p_{\phi_{i}}^2}{\tau^2}-
\phi_{i}^{\prime2}\Big)\,,\nonumber\\
&&\{p_{\phi_i},C[N_g]\}=
(N_g\tau\phi^\prime_i)^\prime\,,\nonumber
\end{eqnarray*}
and
\begin{eqnarray*}
&&\{\gamma,C_\theta[N^\theta_g]\}=
2N_{g}^{\theta\prime}+N^\theta_g\gamma^\prime\,,
\nonumber\\
&&\{\tau,C_\theta[N^\theta_g]\}=N^\theta_g \tau^\prime\,,\nonumber\\
&&\{\phi_i,C_\theta[N^\theta_g]\}=N^\theta_g\phi_i^\prime\,,\nonumber\\
&&\{p_\gamma,C_\theta[N^\theta_g]\}=(N^\theta_g
p_\gamma)^\prime\,,\nonumber\\
&&\{p_\tau,C_\theta[N^\theta_g]\}=(N^\theta_g p_\tau)^\prime\,,\nonumber\\
&&\{p_{\phi_i},C_\theta[N^\theta_g]\}=(N^\theta_g p_{\phi_i})^\prime\,.\nonumber\\
\end{eqnarray*}
A straightforward calculation shows that the constraints are
first class in Dirac terminology, or equivalently that $\Gamma_{c}$
is a coisotropic submanifold of $\Gamma$. Indeed, the Poisson
algebra of the constraints is a proper Lie algebra
\begin{eqnarray*}
&&\{C[N_g],C[M_g]\}=C_\theta[N_g
M^\prime_g-M_g N^\prime_g]\,,\nonumber\\
&&\{C[N_g],C_\theta[N^\theta_g]\}=C[N_g
M^{\theta\prime}_g-M^\theta_g N_g^\prime]\,,\nonumber\\
&&\{C_\theta[N^\theta_g],C_\theta[M^\theta_g]\}=C_\theta[N^\theta_g
M^{\theta\prime}_g-M^\theta_g N^{\theta\prime}_g]\,.\nonumber
\end{eqnarray*}
Notice also that, as a consequence of the introduction of the
suitable exponential factor $e^{\gamma/2}$ in (\ref{ta}) we have
a \textit{closed} gauge algebra \cite{Romano:1996ep} (i.e. with structure
\textit{constants}).

In order to proceed we would like to isolate the true physical
degrees of freedom of the model. As is well known there are several
possible ways to do this. The usual ones are gauge fixing, i.e. the isolation of a single
point per gauge orbit by imposing appropriate extra conditions on
the phase space variables, and phase space reduction
--that requires us to find a way to effectively quotient the phase
space by the equivalence relation loosely defined as ``belonging to
the same orbit''. The successful implementation of the reduction allows us
not only to label gauge orbits but also provides us with important
mathematical structures (topological, symplectic,...) from the ones
present in the initial phase space. Here we will see that a \textit{partial}
gauge fixing (\emph{deparameterization}) can provide us with another
interesting way to deal with the system because it can be described
by a time-dependent, quadratic, Hamiltonian \cite{Misner, Berger:1973, Pierri:2000ri}.
As we will show below this is also true for the other spatial topologies.
If one is interested in quantizing the model one can alternatively use the
Hamiltonian formulation described above to attempt a Dirac quantization.

The
Hamiltonian vector fields associated to the weighted constraints
$C[N_g]$, $C_{\theta}[N_{g}^{\theta}]$ are tangential to
$\Gamma_c$ and define the degenerate directions of the presymplectic form $\omega|_{\Gamma_c}$.
The deparameterization procedure is based on the choice of one of these
Hamiltonian vector fields to define an evolution vector field
$E_{H_R}$, generated by a reduced Hamiltonian $H_R$
of a generically non-autonomous system.
With this aim in mind, we will impose gauge fixing conditions
in such a way that at least one of the first class constraints $\mathcal{C}$ is not fixed. This will be used to define dynamics. Any remaining constraints left over by the (partial) gauge fixing will generate residual gauge symmetries.

Let $\iota:\Gamma_G\rightarrow\Gamma_c$ denote the embedding of the gauge fixed surface given
by the first class constraints and the gauge fixing conditions; the
pull-back of the presymplectic form to this surface, $\iota^{*}\omega$,
has a single degenerate direction defined by  the
Hamiltonian vector field $E_{H_R}$.
Select then a suitable phase space variable $T$ such that
$E_{H_R}(T)=1$. The level surfaces of $T$ are all
diffeomorphic to a manifold $\Gamma_{R}$ and transverse to
$E_{H_R}$, defining a foliation of $\Gamma_{G}$ with $T$
as global time function. In that case,
$\iota^{*}\omega=-\mathrm{d}T\wedge\mathrm{d}H_R+\omega_{R}$,
$E_{H_R}=\partial_{T}+X_{H_R}$, where $\omega_{R}$ is a weakly
non-degenerate form, and $(\Gamma_R,\omega_R,H_R(T))$ define a
non-autonomous Hamiltonian system. Any remaining first class constraints will define a constrain
 submanifold in $\Gamma_{R}$.

The conditions that are usually considered for this problem
\cite{Pierri:2000ri,Romano:1996ep,Corichi:2002vy,Torre:2002xt} are
\begin{eqnarray}
&&\tau^\prime=0\,,\label{gauge1}\\
&&p_\gamma^\prime=0\,.\label{gauge2}
\end{eqnarray}
They mean that both $\tau$ and $p_\gamma$ take the
same value irrespective of the point of $\mathbb{S}^1$ but they do not specify which
one. Notice that conditions of the type $\tau=t_0$ or $p_\gamma=-p$ with
$t_0,p\in\mathbb{R}$ not only would tell us that $\tau$ and
$p_\gamma$ are independent of $\theta$ but also assign a fixed value
to them, thus removing additional degrees of freedom.\footnote{A
useful example to appreciate the difference between taking some
derivatives to be zero and fixing the values of the functions is to
consider the straight line $x_1=x_2=x_3$ in $\mathbb{R}^3$ where all
the points have equal coordinates in contrast with the point
$x_1=x_2=x_3=1$.}

This means that when using (\ref{gauge1},\ref{gauge2}) there is
still a dynamical mode in $\tau$ that may vary in the evolution --at
the end of the day it will be identified with the time parameter--
but is constant on every spatial slice in the 3+1 decomposition. The
fact that this class of models have an initial spacetime singularity
suggests that there are interesting interpretive issues as far as
the equivalence of the different choices of gauge fixing is
concerned (how does this singularity manifest itself after a
\textit{full} gauge fixing? how does it show up if other gauge
fixing conditions are used?). It should also be pointed out that although it is
possible to think of the condition $\tau^\prime=0$ as a one
parameter family of gauges $\tau=t$, with $t\in(0,\infty)$, it is
dangerous to use it in this last form when computing Poisson
brackets (it would be something like ``mixing parametric and
implicit equations'') as can be checked by explicit computations. In
this case the correct attitude would be to work in the extended
(``odd-dimensional'') phase space, mathematically described as a
cosymplectic or contact manifold, incorporating a time variable and
employ the usual techniques for non-autonomous Hamiltonian systems
\cite{Leon1,Leon2}.

A convenient way to discuss gauge fixings is to describe our family
of gauge conditions by introducing an orthonormal basis of weight
functions on $\mathbb{S}^1$
$$
Y_n(\theta):=\frac{1}{\sqrt{2\pi}}e^{in\theta}\,,\,\,\,n\in\mathbb{Z}\,,
$$
and consider the family of constraints $C[Y_{n}]$,
$C_\theta[Y_{n}]$. By expanding now
\begin{eqnarray*}
&&\tau=\sum_{n\in\mathbb{Z}}\tau_{n}Y_{n}\,, \quad
p_{\gamma}=\sum_{n\in\mathbb{Z}}p_{\gamma_n}Y_{n}\,,
\end{eqnarray*}
with
\begin{eqnarray*}
&&\tau_n=\frac{1}{\sqrt{2\pi}}\int_{\mathbb{S}^1}e^{-in\theta}\tau\,,\quad
p_{\gamma_n}=\frac{1}{\sqrt{2\pi}}\int_{\mathbb{S}^1}e^{-in\theta}p_{\gamma}\,,
\end{eqnarray*}
the previous gauge fixing conditions become
\begin{equation}
\tau_n=0\,,\quad p_{\gamma_n}=0\,,\quad \forall\,n\in
\mathbb{Z}-\{0\}\,.
\end{equation}
In order to see if this is a good gauge fixing (and, alternatively,
find out if some gauge freedom is left) we compute
\begin{eqnarray}
&&\{\tau_n,C[Y_m]\}
\approx-\frac{1}{\sqrt{2\pi}}\delta_{nm}p_{\gamma_0}\,,\quad
\{\tau_n,C_\theta[Y_m]\}\approx0\,,\nonumber\\
&&\{p_{\gamma_n},C[Y_m]\}\approx0\,,\quad \{p_{\gamma
n},C_\theta[Y_m]\}\approx\frac{in}{\sqrt{2\pi}}\delta_{nm}p_{\gamma_0}\,,\nonumber
\end{eqnarray}
where $n\in\mathbb{Z}-\{0\}$, $m\in\mathbb{Z}$, and the symbol
$\approx$ denotes equality on the hypersurface defined by the gauge
fixing conditions and the constraints, the so-called gauge fixing
surface $\Gamma_{G}\subset\Gamma_{c}$. Notice that with this way of
writing the constraints (without the extra terms that would be
present if we had not introduced the exponential prefactor in
(\ref{ta})) the gauge transformations of $\tau$ and $p_\gamma$ only
involve these objects themselves. It is convenient to write the
previous expressions in table form
\begin{center}
\begin{tabular}{|c|c c|c c|c|}
\hline\hline
&$\tau_1=0$&$p_{\gamma_1}=0$&$\tau_{-1}=0$&$p_{\gamma_{-1}}=0$&$\dots$\\\hline
$C[Y_{0}]$&$0$&$0$&$0$&$0$&$\dots$\\
$C_\theta[Y_0]$&$0$&$0$&$0$&$0$&$\dots$\\\hline
$C[Y_{1}]$&$-\frac{1}{\sqrt{2\pi}}p_{\gamma0}$&$0$&$0$&$0$&$\dots$\\
$C_\theta[Y_1]$&$0$&$\frac{i}{\sqrt{2\pi}}p_{\gamma0}$&$0$&$0$&$\dots$\\\hline
$C[Y_{-1}]$&$0$&$0$&$-\frac{1}{\sqrt{2\pi}}p_{\gamma0}$&$0$&$\dots$\\
$C_\theta[Y_{-1}]$&$0$&$0$&$0$&$-\frac{i}{\sqrt{2\pi}}p_{\gamma0}$&$\dots$\\\hline
$\vdots$&$\vdots$&$\vdots$&$\vdots$&$\vdots$&$\ddots$\\
\hline\hline
\end{tabular}
\end{center}
As we can see the only constraints that are not gauge-fixed by the
conditions introduced above, as long as $p_{\gamma_0}\neq0$, are
$C[1]=0$ and
$C_\theta[1]=0$. From now on we will
consider the sector $p_{\gamma_0}<0$. As we can see we have two
first class constraints left over by our partial gauge fixing
\begin{eqnarray}
&&\int_{\mathbb{S}^1}
\bigg((2\tau^{\prime\prime}-\gamma^\prime\tau^\prime)-p_\gamma
p_\tau+\frac{1}{2}\sum_{i}\Big(\frac{p_{\phi_{i}}^2}{\tau}
+\tau\phi_{i}^{\prime2}\Big)\bigg)\approx0\,,\label{const_res_N}\\
&&\int_{\mathbb{S}^1} \Big(-2p_\gamma^\prime+p_\gamma
\gamma^\prime+p_\tau \tau^\prime+\sum_{i}p_{\phi_{i}}\phi_{i}^\prime
\Big)\approx0\,.\label{const_res_N_theta}
\end{eqnarray}
We can pullback the relevant geometric objects to the submanifold
$\Gamma_{G}$ defined by the gauge fixing conditions to eliminate
some of the variables in our model. Denoting by
$\iota:\Gamma_{G}\rightarrow\Gamma$ the immersion map, the pullback of
the (weakly) symplectic form (\ref{ws2f}) becomes
\begin{eqnarray}\label{GG2-f}
\iota^{*}\omega=\mathrm{d}\gamma_0\wedge \mathrm{d}
p_{\gamma_0}+\mathrm{d}\tau_0\wedge \mathrm{d}
p_{\tau_0}+\sum_{i}\int_{\mathbb{S}^1} \delta\phi_i\wedge \delta
p_{\phi_i}\,.\label{symplectic1}
\end{eqnarray}
The pullback of the constraints
(\ref{const_res_N},\ref{const_res_N_theta}) is
\begin{eqnarray}
&&\mathcal{C}:=-p_{\gamma_0} p_{\tau_0}+\frac{1}{2}\sum_{i}
\int_{\mathbb{S}^1}\left(\sqrt{2\pi}\frac{p^2_{\phi_i}}{\tau_0}+
\frac{\tau_0}{\sqrt{2\pi}}\phi^{\prime2}_i\right)\approx0\,,\label{Cgf1}\\
&&\mathcal{C}_\theta:=\sum_{i}\int_{\mathbb{S}^1}p_{\phi_i}\phi^\prime_i\approx0\,.\label{Cgf2}
\end{eqnarray}
Let us look now at the gauge transformations of $\tau_0$ generated
by (\ref{Cgf1})
\begin{eqnarray}
&&\{\tau_0,\mathcal{C}\}\approx-p_{\gamma_0}\,,\nonumber\\
&&\{p_{\gamma_0},\mathcal{C}\}\approx0\,.\nonumber
\end{eqnarray}
If $s\in (0,\infty)$ parameterizes the gauge orbits we see that on
them we have $\tau_0=ps$ and $p_{\gamma_0}=-p$, with $p>0$. This
suggests that a simplification of our model will occur if we
introduce a canonical transformation where $\tau_0$ and
$p_{\gamma_0}$ are substituted for new canonical variables. Indeed,
the canonical transformation \cite{Cortez}
\begin{eqnarray}
&\displaystyle\tau_0=TP\,,\hspace{2.3cm}&p_{\tau_0}=\frac{p_T}{P}\,,\nonumber\\
&\displaystyle\gamma_0=-\frac{1}{\sqrt{2\pi}}\big(Q+\frac{p_T}{P}T\big)\,,
&p_{\gamma_0}=-\sqrt{2\pi}P\,,\nonumber
\end{eqnarray}
with $(Q,P>0)$, and $(T,p_T)$ canonically conjugate pairs, allows us
to write
\begin{equation}
\mathcal{C}=p_T+\frac{1}{2}\sum_{i}\int_{\mathbb{S}^1}\left(\frac{p^2_{\phi_i}}{PT}+
PT\phi^{\prime2}_i\right)\approx0\,.\label{Cgf3}
\end{equation}
Finally the canonical transformation (here $(\tilde{Q},\tilde{P})$
and $(\varphi_i,p_{\varphi_i})$ are new canonical pairs
\cite{Cortez})
\begin{eqnarray}
&&\tilde{Q}:=PQ+\frac{1}{2}\sum_{i}\int_{\mathbb{S}^1}p_{\phi_i}\phi_i\,,
\quad \tilde{P}:=\log P\,,\nonumber\\
&&\varphi_i:=\sqrt{P}\phi_i\,,\quad
p_{\varphi_i}:=\frac{1}{\sqrt{P}}p_{\phi_i}\,,
\end{eqnarray}
turns the constraints (\ref{Cgf3},\ref{Cgf2}) into
\begin{eqnarray}
&&\mathcal{C}=p_T+\frac{1}{2}\sum_{i}\int_{\mathbb{S}^1}\bigg(\frac{p^2_{\varphi_i}}{T}+
T\varphi^{\prime2}_i\bigg)\approx0\,,\label{C1}\\
&&\mathcal{C}_\theta=\sum_{i}\int_{\mathbb{S}^1}p_{\varphi_i}\varphi^\prime_i\approx0\label{C2}\,,
\end{eqnarray}
and the 2-form (\ref{GG2-f}) becomes
\begin{equation}\label{omegaHR}
\iota^{*}\omega=\mathrm{d}\tilde{Q}\wedge\mathrm{d}\tilde{P}+\sum_{i}\int_{\mathbb{S}^{1}}
\delta\varphi_{i}\wedge\delta
p_{\varphi_i}+\mathrm{d}T\wedge\mathrm{d}p_T\,.
\end{equation}
The fact that (\ref{C1}) is linear in $p_T$ allows us to interpret
the 4-tuple
$((0,\infty)\times\Gamma_{R},\mathrm{d}t,\omega_R,H_{R})$ as a
non-autonomous Hamiltonian system with $T=t$ as the time parameter,
restricted to verify the global constraint (\ref{C2}). The reduced
phase space $\Gamma_{R}$ is coordinatized now by the canonical pairs
$(\tilde{Q},\tilde{P};\varphi_{i},p_{\varphi_i})$ and is endowed
with the (weakly) symplectic form
\begin{equation}\label{omegaR}
\omega_{R}:=\mathrm{d}\tilde{Q}\wedge\mathrm{d}\tilde{P}+\sum_{i}\int_{\mathbb{S}^{1}}
\delta\varphi_{i}\wedge\delta p_{\varphi_i}\,.
\end{equation}
The reduced time-dependent Hamiltonian
$H_{R}(t):\Gamma_{R}\rightarrow\mathbb{R}$ is given by
\begin{equation}\label{HR}
H_{R}(t)=\frac{1}{2}\sum_{i}\int_{\mathbb{S}^1}\bigg(\frac{p^2_{\varphi_i}}{t}+
t\varphi^{\prime2}_i\bigg)\,,
\end{equation}
and the evolution vector field is given by
$$
E_{H_R}=\frac{\partial}{\partial
t}+\sum_{i}\int_{\mathbb{S}^{1}}\left(\frac{p_{\varphi_i}}{t}\frac{\delta}{\delta\varphi_i}+t\varphi_{i}^{\prime\prime}\frac{\delta}{\delta
p_{\varphi_i}}\right).
$$
This defines the only degenerate direction of (\ref{omegaHR}).
Although the form of the Hamiltonian that we
have just obtained seems to suggest that the gravitational and matter scalars are not coupled, in
fact the constraint (\ref{C2}) shows that this is not the
case\footnote{The matter fields act as sources for the gravitational
field, hence, the solutions to the Einstein equations should depend
on the matter content.}. Notice also that the canonical pair
$(\tilde{Q},\tilde{P})$ describes a global degree of freedom even though they are
constants of motion under the dynamics generated by (\ref{HR}). The singularities that must be present in this case as a consequence of the Hawking-Penrose theorems \cite{Wald} can be understood as coming from the singular behavior at $t=0$ of the Hamiltonian (\ref{HR}).

Finally, it is possible to recover the original 4-dimensional
spacetime from this 3-dimensional formulation. First notice that the
gauge fixing conditions defining the deparameterization are
preserved under the dynamics if and only if the lapse and shift
functions $N$ and $N^{\theta}$ are constant. By redefining the
coordinate $\theta$ as in \cite{Mena:1997} we can eliminate the
shift function from the metric. We can proceed in an analogous way
for the lapse function to make it equal to 1. Once we integrate the
Hamiltonian equations corresponding to (\ref{HR}), undo the
canonical transformation defined above, and solve the constraint
$C_{\theta}=0$ in order to obtain the $\gamma$ function, we uniquely
determine the 3-metric (\ref{metric}), and hence the original
4-metric.

\section{$\mathbb{S}^1\times\mathbb{S}^2$ Gowdy models coupled to massless
scalars}{\label{handle}}

Let us consider now the three-handle
${^{\tres}}\Sigma=\mathbb{S}^1\times\mathbb{S}^2$, parameterized as
$(e^{i\xi},e^{i\sigma}\sin\theta,\cos\theta)$ with
$\theta\in[0,\pi]$, $\xi,\sigma\in\mathbb{R}(\mathrm{mod}\,2\pi)$.
Using the group parametrization introduced above we can write the
$G^\dos$-group action in the form
$$
(g_1,g_2)\cdot(e^{i\xi},e^{i\sigma}\sin\theta,\cos\theta)=
(e^{x_1},e^{x_2})\cdot(e^{i\xi},e^{i\sigma}\sin\theta,\cos\theta)=
(e^{i(x_1+\xi)},e^{i(x_2+\sigma)}\sin\theta,\cos\theta)\,.
$$
The action of the two $U(1)$ subgroup factors of $G^\dos$ is
\begin{eqnarray*}
&&(1,e^{ix})\cdot(e^{i\xi},e^{i\sigma}\sin\theta,\cos\theta)=
(e^{i\xi}, e^{i(x+\sigma)}\sin\theta,\cos\theta)\,,\nonumber\\
&&(e^{ix},1)\cdot(e^{i\xi},e^{i\sigma}\sin\theta,\cos\theta)=
(e^{i(x+\xi)},e^{i\sigma}\sin\theta,\cos\theta)\,.\nonumber
\end{eqnarray*}
The corresponding tangent vectors at each point of
${^{\tres}}\Sigma$, obtained by differentiating the previous
expressions with respect to $x$ at $x=0$, are
\begin{eqnarray*}
&&(0,ie^{i\sigma}\sin\theta,0)\,,\nonumber\\
&&(ie^{i\xi},0,0)\,.\nonumber
\end{eqnarray*}
As we can see the second one is never zero but the first one
vanishes at the poles of the sphere $\mathbb{S}^2$ where
$\theta=0,\pi$. This corresponds to the circumferences given by
$(e^{i\xi},0,1)$ and $(e^{i\xi},0,-1)$. It is straightforward to
verify that both fields commute. In view of all this we perform a Geroch reduction
by using the non-vanishing Killing. After a
suitable conformal transformation the field equations can be derived
from an action of the form (\ref{actr}) with
$^{\scriptscriptstyle{\dos}}\Sigma=\mathbb{S}^2$. All the fields in this action are
defined on $\mathbb{S}^2$ and are symmetric under the symmetries generated by the
remaining Killing $\sigma^a$. Since this Killing vector vanishes at
the poles of the sphere $\mathbb{S}^2$ we cannot build an everywhere
orthonormal basis that involves this vector. In fact, we know that
as $\mathbb{S}^2$ is not paralelizable this is impossible on general
grounds. We nevertheless will consider the triplet of vectors
$(n^a,\hat{\theta}^a,\sigma^a/\tau)$ whenever it is different from zero
(for all $\theta\neq0,\pi$). Taking again the definition (\ref{ta}),
the form of the metric is the same as in the $\mathbb{T}^3$ case
(\ref{metric}). The symmetry of the problem implies also that
$(N,N^\theta,\gamma,\tau,\phi_{i})$  are constant on the orbits of
the Killing field $\sigma^a$.

A very important issue now is the regularity of the metric. From a
classical point of view the final outcome of the Hamiltonian
analysis of the system is a set of equations whose solutions allow
us to reconstruct a four dimensional spacetime metric and a set of
scalar fields satisfying the coupled Einstein-Klein Gordon
equations. This means that once we decide the functional space to
which this metric belongs this will imply that the objects that
appear during the dimensional reduction, gauge fixing and so on may
be subject to some regularity conditions. In the $\mathbb{T}^3$ case
these are simple smoothness requirements but in the present case,
due to the existence of a symmetry axis, these are more complicated.
The regularity conditions that the metric components for an axially
symmetric metric must verify can be deduced as in
\cite{Chrusciel:1990zx,Rinne:2005sk}. By using the coordinates
$(t,\theta,\sigma,\xi)$, we can write the original 4-metric
${^{\cuatro}}g_{ab}$ as
\begin{eqnarray}
{^{\cuatro}}g_{ab}&=&e^{(\gamma-\phi_1)}[(N^{\theta
2}-N^{2})(\mathrm{d}t)_a(\mathrm{d}t)_b+
2N^{\theta}(\mathrm{d}t)_{(a}(\mathrm{d}\theta)_{b)}+(\mathrm{d}\theta)_a(\mathrm{d}\theta)_b]
\label{4metric}\\
&+& \tau^{2}e^{-\phi_1}(\mathrm{d}\sigma)_a(\mathrm{d}\sigma)_b+
e^{\phi_1}(\mathrm{d}\xi)_a(\mathrm{d}\xi)_b\,.\nonumber
\end{eqnarray}
This means that we have the following regularity conditions for
$\theta\rightarrow0,\pi$ (if we
impose analyticity, otherwise we need only to know the asymptotic
behavior for small values of $\sin{\theta}$)
\begin{eqnarray}
&&e^{(\gamma-\phi_1)}(N^{\theta 2}-N^2)=A(t,\cos\theta)\,,\label{reg1}\\
&&e^{(\gamma-\phi_1)}N^\theta=B(t,\cos\theta)\sin\theta\,, \label{reg2}\\
&&e^{\phi_1}=C(t,\cos\theta)\,,\label{reg3}\\
&&e^{\gamma-\phi_1}=D(t,\cos\theta)+E(t,\cos\theta)\sin^2\theta\,, \label{reg4}\\
&&\tau^2e^{-\phi_1}=\sin^2\theta[D(t,\cos\theta)-E(t,\cos\theta)\sin^2\theta
]\,,\label{reg5}
\end{eqnarray}
where $A,\,B,\,C,\,D,\,E$ are analytic in their arguments (despite
of the fact that they also depend on $t$, as we will use them in the
Hamiltonian formulation of the model we will not write the $t$
dependence explicitly in the following). Notice, in particular, that
the functions $D$ and $E$ both appear in the last two expressions.
The conditions for the fields themselves (dropping the $t$
dependence) are
\begin{eqnarray}
&&\phi_i=\hat{\phi}_{i}(\cos\theta)\,,\label{reg_1}\\
&&\gamma=\hat{\gamma}(\cos\theta)\,,\label{reg_2}\\
&&N^\theta=\hat{N}^{\theta}(\cos\theta)\sin\theta\,,\label{reg_3}\\
&&N=\hat{N}(\cos\theta)\,,\label{reg_4}\\
&&\tau=\hat{T}(\cos\theta)\sin\theta\,,\label{reg_5}\\
&&\tau^2e^{-\gamma}=\frac{D(\cos\theta)-E(\cos\theta)\sin^2\theta}{D(\cos\theta)
+E(\cos\theta)\sin^2\theta}\sin^2\theta\,,\label{reg_6}
\end{eqnarray}
where
$\hat{\phi}_{i},\hat{\gamma},\hat{N}^{\theta},\hat{N},\hat{T}:[-1,1]\rightarrow\mathbb{R}$
($\hat{N}>0$) and can be written as functions of
$A,\,B,\,C,\,D,\,E$. They must be differentiable in $(-1,1)$ and
their right and left derivatives at $\pm1$ must be defined
(equivalently they must be $\mathcal{C}^\infty$ in $(-1,1)$ with
bounded derivative). Several comments are in order now. First we
have been able to write all the relevant fields in such a way that
their singular dependence has been factored out ($\sin\theta$
\textit{is not} a smooth function on the sphere). The functions
defined on $\mathbb{S}^2$ as
$\hat{\phi}_{i}\circ\cos\theta,\hat{\gamma}\circ\cos\theta,
\hat{N}^{\theta}\circ\cos\theta,\hat{N}\circ\cos\theta,\hat{T}\circ\cos\theta$
can be alternatively viewed as analytic functions on the sphere
invariant under rotations around its symmetry axis and can be
considered as the basic fields to describe our system. In fact we
will do so in the following. We will refer to these functions on the
sphere as
$\hat{\phi}_{i},\hat{\gamma},\hat{N}^{\theta},\hat{N},\hat{T}$
(without the $\circ\cos\theta$ that will only be used if the
possibility of confusion arises) and collectively as the
\emph{hat}-fields. In the following we will write everything in
terms of them. Second we can see that condition (\ref{reg_6}) implies that
\begin{equation}
\hat{T}(\pm1)=e^{\hat{\gamma}(\pm1)/2}\,.\label{polarS2}
\end{equation}
This means that the values of the fields $\hat{T}$ and
$\hat{\gamma}$ at the poles of the sphere are not independent of
each other. This is a new feature, not present for the
$\mathbb{T}^3$ topology, that must be taken into account. As we will
see these are necessary ingredients to guarantee the consistency of
the model.

Given a smooth (and axially symmetric) function on
$\mathbb{S}^2$ its Lie derivative along $\theta^a$,
$\mathcal{L}_\theta$, cannot necessarily be extended as a smooth
function on the sphere. The function $\cos\theta$ itself is an
example of this because $\mathcal{L}_\theta\cos\theta=-\sin\theta$.
We can, however, define a smooth derivative $f^\prime$ for a smooth
axially symmetric function as the extension of
$f^\prime:=-\frac{1}{\sin\theta}\partial_\theta f$ to $\mathbb{S}^2$
(this is formally done by considering $f$ as a function of
$\cos\theta$ and differentiating). In the following the
\textit{prime} symbol will always refer to this derivative.

It is natural to consider in this case ${^{\dos}}\mathrm{e}$ as the
fiducial 2-form associated to a round metric, such that
${^{\dos}}\mathrm{e}_{ab}\theta^{a}\sigma^{b}=\hat{N}e^{\hat{\gamma}}\hat{T}\sin\theta/|g|^{1/2}=\sin\theta$,
i.e.
${^{\dos}}\mathrm{e}=\sin\theta\mathrm{d}\theta\wedge\mathrm{d}\sigma$.
Taking this into account we get an action
\begin{eqnarray}
&&\frac{1}{16\pi G_3}\int_{t_0}^{t_1}\!\!\!\!\!
\mathrm{d}t\int_{\mathbb{S}^2}\!\!\!^{\dos}\mathrm{e}\bigg[
\hat{N}[(\hat{\gamma}^\prime\hat{T}^\prime-2\hat{T}^{\prime\prime})\sin^2\theta+
(6\hat{T}^{\prime}-\hat{\gamma}^\prime\hat{T})\cos\theta+2\hat{T}]\nonumber\\
&&\hspace{30mm}+\frac{1}{\hat{N}}[\hat{N}^\theta
\hat{T}\cos\theta-\dot{\hat{T}}-\hat{N}^\theta
\hat{T}^\prime\sin^2\theta]\,[\dot{\hat{\gamma}}
+(2\hat{N}^{\theta\prime}+\hat{N}^\theta\hat{\gamma}^\prime)\sin^2\theta-2\hat{N}^\theta\cos\theta]
\nonumber\\
&&\hspace{30mm}+\frac{\hat{T}}{2\hat{N}}\sum_{i}\bigg({\dot{\hat{\phi}}}\,^2_i
+2\hat{N}^\theta\dot{\hat{\phi}}_i\,\hat{\phi}_i^\prime\sin^2\theta+(\hat{N}^{\theta2}\sin^2\theta
-\hat{N}^2)\hat{\phi}_i^{\prime2}\sin^2\theta\bigg)\bigg]\,.\nonumber
\end{eqnarray}
As we can see it is expressed as the integral of a smooth
function on the sphere. This is so because all the fields that
appear in the integrand are either the $hat$-fields, their prime
derivatives or smooth functions of $\cos\theta$. The Hamiltonian can
be readily derived from the previous action and, as in the
$\mathbb{T}^{3}$ case, is of the form
$H=C[\hat{N}]+C_\theta[\hat{N}^\theta]$ with
\begin{eqnarray}
C[\hat{N}]&=&\int_{\mathbb{S}^2}{^{\dos}}\mathrm{e}\hat{N}C\label{C1S2}\\
&:=&\int_{\mathbb{S}^2}\!\!\!^{\dos}\mathrm{e}\hat{N}\bigg[ -16\pi
G_3 p_{\hat{\gamma}}p_{\hat{T}}+\frac{1}{16\pi
G_3}\big[(2\hat{T}^{\prime\prime}-\hat{\gamma}^\prime\hat{T}^\prime)\sin^2\theta
+(\hat{\gamma}^\prime\hat{T}-6\hat{T}^\prime)\cos\theta-2\hat{T}\big]\nonumber\\
&&\hspace{1cm}+\frac{1}{2}\sum_{i}\bigg(\frac{16\pi
G_3}{\hat{T}}p_{\hat{\phi}_i}^2+\frac{\hat{T}}{16\pi
G_3}\hat{\phi}_i^{\prime2}\sin^2\theta\bigg)\bigg],\nonumber\\
C_\theta[\hat{N}^\theta]&=&\int_{\mathbb{S}^2}{^{\dos}}\mathrm{e}\hat{N}^{\theta}C_{\theta}\label{C2S2}
\\&:=&\int_{\mathbb{S}^2}\!\!\!^{\dos}\mathrm{e}\hat{N}^\theta
\bigg([2p^\prime_{\hat{\gamma}}-\hat{\gamma}^\prime
p_{\hat{\gamma}}- \hat{T}^\prime
p_{\hat{T}}-\sum_{i}\hat{\phi}_i^\prime
p_{\hat{\phi}_i}]\sin^2\theta +[\hat{T}
p_{\hat{T}}-2p_{\hat{\gamma}}]\cos\theta\bigg).\nonumber
\end{eqnarray}
Again, the dynamical variables are restricted to belong to a
constraint surface $\Gamma_{c}\subset\Gamma$ in the canonical phase
space of the system $(\Gamma,\omega)$, globally defined by the
constraints $C=0$, $C_{\theta}=0$. $\Gamma$ is coordinatized by the
conjugated pairs
$(\hat{\gamma},p_{\hat{\gamma}};\hat{T},p_{\hat{T}};\hat{\phi}_{i},p_{\hat{\phi}_i})$
and endowed with the standard (weakly) symplectic form
\begin{equation}\label{omegaS2XS1}
\omega:=\int_{\mathbb{S}^2}{^{\dos}}\mathrm{e}\bigg(\delta\hat{\gamma}\wedge\delta
p_{\hat{\gamma}}+\delta\hat{T}\wedge\delta
p_{\hat{T}}+\sum_{i}\delta\hat{\phi}_i\wedge\delta
p_{\hat{\phi}_i}\bigg)\,.
\end{equation}
The gauge transformations generated by these constraints
are\footnote{In the following $16\pi G_3=1$.}
\begin{eqnarray*}
&&\{\hat{\gamma},C[\hat{N}_g]\}=-\hat{N}_gp_{\hat{T}}\,,
\nonumber\\
&&\{\hat{T},C[\hat{N}_g]\}=-\hat{N}_g p_{\hat{\gamma}}\,,\nonumber\\
&&\{\hat{\phi_i},C[\hat{N}_{g}]\}=\frac{\hat{N}^g}{\hat{T}}p_{\hat{\phi}_i}\,,\nonumber\\
&&\{p_{\hat{\gamma}},C[\hat{N}_g]\}= \hat{N}_{g}^{\prime}(\hat{T}\cos\theta-\hat{T}^\prime\sin^2\theta)+
\hat{N}_g(\hat{T}+3\hat{T}^\prime\cos\theta-\hat{T}^{\prime\prime}\sin^2\theta),\nonumber\\
&&\{p_{\hat{T}},C[\hat{N}_{g}]\}=\hat{N}_{g}^{\prime}(2\cos\theta-\hat{\gamma}^\prime\sin^2\theta)+
\hat{N}_{g}(\hat{\gamma}^\prime\cos\theta-\hat{\gamma}^{\prime\prime}\sin^2\theta)-2\hat{N}_{g}^{\prime\prime}
\sin^2\theta\nonumber\\
&&\hspace{2.6cm}+\frac{\hat{N}_g}{2}\sum_{i}\bigg(\frac{p_{\hat{\phi}_i}^2}{\hat{T}^2}-
\sin^2\theta\hat{\phi}_i^{\prime2}\bigg)\,,\nonumber\\
&&\{p_{\hat{\phi}_i},C[\hat{N}_g]\}=
\hat{N}_{g}^{\prime}\hat{T}\hat{\phi}_i^{\prime}\sin^2\theta+\hat{N}_{g}[(\hat{T}^\prime\hat{\phi}_i^\prime+
\hat{T}\hat{\phi}_i^{\prime\prime})\sin^2\theta-2\hat{T}\hat{\phi_i^\prime}\cos\theta]\,,\nonumber\\
\end{eqnarray*}
and
\begin{eqnarray*}
&&\{\hat{\gamma},C_\theta[\hat{N}_{g}^\theta]\}=-2\hat{N}_{g}^{\theta\prime}\sin^2\theta+
\hat{N}_{g}^\theta(2\cos\theta-\hat{\gamma}^\prime\sin^2\theta)\,,\nonumber\\
&&\{\hat{T},C_\theta[\hat{N}_{g}^\theta]\}=\hat{N}_{g}^\theta(\hat{T}\cos\theta-\hat{T}^\prime\sin^2\theta)\,,\nonumber\\
&&\{\hat{\phi}_i,C_\theta[\hat{N}_{g}^\theta]\}=-\hat{N}_{g}^\theta\hat{\phi}_i^\prime\sin^2\theta\,,\nonumber\\
&&\{p_{\hat{\gamma}},C_\theta[\hat{N}_{g}^\theta]\}=
\hat{N}_{g}^\theta(2p_{\hat{\gamma}}\cos\theta-p^\prime_{\hat{\gamma}}\sin^2\theta)-
\hat{N}_{g}^{\theta\prime}p_{\hat{\gamma}}\sin^2\theta\,,\nonumber\\
&&\{p_{\hat{T}},C_\theta[\hat{N}_{g}^\theta]\}=\hat{N}_{g}^\theta(p_{\hat{T}}\cos\theta-p^\prime_{\hat{T}}\sin^2\theta)-
\hat{N}_{g}^{\theta\prime}p_{\hat{T}}\sin^2\theta\,,\nonumber\\
&&\{p_{\hat{\phi}_i},C_\theta[\hat{N}_{g}^\theta]\}=\hat{N}_{g}^\theta(2p_{\hat{\phi}_i}\cos\theta-
p^\prime_{\hat{\phi}_i}\sin^2\theta)-\hat{N}_{g}^{\theta\prime}p_{\hat{\phi}_i}\sin^2\theta\,.\nonumber\\
\end{eqnarray*}
As is the $\mathbb{T}^{3}$ case, $\Gamma_c\subset\Gamma$ is a first
class submanifold as can be seen by computing the Poisson algebra of
the (weighted) constraints
\begin{eqnarray*}
&&\hspace{-2mm}\{C[\hat{N}_g],C[\hat{M}_g]\}=C_\theta[\hat{M}_g
\hat{N}_g^{\prime}-\hat{N}_g
\hat{M}_g^{\prime}]\,,\nonumber\\
&&\hspace{-2mm}\{C[\hat{N}_g],C_\theta[\hat{M}_g^\theta]\}=C[(\hat{M}_{g}^\theta
\hat{N}_{g}^{\prime}-\hat{N}_g
\hat{M}_{g}^{\theta\prime})\sin^2\theta+\hat{N}_g\hat{M}_{g}^\theta\cos\theta]\,,\nonumber\\
&&\hspace{-2mm}\{C_\theta[\hat{N}_{g}^\theta],C_\theta[\hat{M}_{g}^\theta]\}=C_\theta[(\hat{M}_{g}^\theta
\hat{N}_{g}^{\theta\prime}-\hat{N}_{g}^\theta
\hat{M}_{g}^{\theta\prime})\sin^2\theta]\,.\nonumber
\end{eqnarray*}
We want to check now the stability of the ``polar constraints''
$(\hat{T}e^{-\hat{\gamma}/2})(\pm1)=1$. To this end we compute
\begin{eqnarray}
&&\hspace{-4mm}\{\hat{T}e^{-\hat{\gamma}/2},C[\hat{N}_{g}]\}=8\pi
G_3\hat{N}_{g}e^{-\hat{\gamma}/2}
\left(\hat{T}p_{\hat{T}}-2p_{\hat{\gamma}}\right),\nonumber\\
&&\hspace{-4mm}\{\hat{T}e^{-\hat{\gamma}/2},C_\theta[\hat{N}_{g}^\theta]\}=
e^{-\hat{\gamma}/2}\big[\hat{T}\hat{N}_g^{\theta\prime}+
\hat{N}_{g}^{\theta}(\frac{1}{2}\hat{T}\hat{\gamma}^\prime-\hat{T}^\prime)\big]\sin^2\theta\,.
\nonumber
\end{eqnarray}
The first expression vanishes at the poles as a consequence of the
constraint (\ref{C2S2}) for $\theta=0,\,\pi$ ($\sin\theta=0$ and
$|\cos\theta|=1$) whereas the second vanishes because of the
$\sin^2\theta$ factor. We then conclude that there are no secondary
constraints coming from the stability of the polar constraints. An
interesting point to highlight here is the fact that these polar
constraints are necessary conditions for the differentiability of
the constraints (\ref{C1S2},\ref{C2S2}).

Deparameterization in this case is carried out by basically following
the same steps as in the $\mathbb{T}^3$ case. Again, in view of the
gauge transformations, we begin by choosing gauge fixing conditions similar
to (\ref{gauge1},\ref{gauge2})
\begin{eqnarray}
&&\hat{T}^\prime=0\,,\label{gauge1S2}\\
&&p_{\hat{\gamma}}^\prime=0\,.\label{gauge2S2}
\end{eqnarray}
We introduce now an orthonormal basis of functions on the subspace
of axially symmetric functions on $\mathbb{S}^2$
$$
Y_{n}(\theta)=
\left({\frac{2n+1}{4\pi}}\right)^{1/2}P_n(\cos\theta)\,,\,\,\,n\in\mathbb{N}\cup\{0\}\,,
$$
where $P_n$ are the Legendre polynomials. By expanding now
\begin{eqnarray*}
\quad\hat{T}=\sum_{n=0}^\infty \hat{T}_nY_{n}\,,\quad
p_{\hat{\gamma}}=\sum_{n=0}^\infty p_{\hat{\gamma}_n}Y_n\,,
\end{eqnarray*}
with
\begin{eqnarray*}
&&\hat{T}_n=\left({\frac{2n+1}{4\pi}}\right)^{1/2}\int_{\mathbb{S}^2}\!\!\!^{\dos}\mathrm{e}
P_{n}(\cos\theta)\hat{T}\,,\quad
p_{\hat{\gamma}_n}=\left({\frac{2n+1}{4\pi}}\right)^{1/2}\int_{\mathbb{S}^2}\!\!\!^{\dos}\mathrm{e}
P_n(\cos\theta)p_{\hat{\gamma}}\,,
\end{eqnarray*}
the conditions (\ref{gauge1S2},\ref{gauge2S2}) become
\begin{equation}
\hat{T}_n=0=p_{\hat{\gamma}_n}\,,\,\,\, \forall\,n\in
\mathbb{N}\,.\label{gauge3S2}
\end{equation}
In order to see if this is a good gauge fixing (and, alternatively,
find out if some gauge freedom is left) we compute
\begin{eqnarray}
&&\{\hat{T}_n,C[Y_m]\}\approx-\sqrt{\frac{(2n+1)(2m+1)}{(4\pi)^3}}p_{\hat{\gamma}0}\int_{\mathbb{S}^2}\!\!\!^{\dos}\mathrm{e}P_n(\cos\theta)P_m(\cos\theta)=\frac{1}{\sqrt{4\pi}}p_{\hat{\gamma}0}\delta_{nm}
\nonumber\\
&&\{\hat{T}_n,C_\theta[Y_m]\}\approx\frac{\hat{T}_0}{4\pi}\sqrt{\frac{(2n+1)(2m+1)}{4\pi}}
\int_{\mathbb{S}^2}\!\!\!^{\dos}\mathrm{e}\cos\theta P_n(\cos\theta)P_m(\cos\theta)\nonumber\\
&&\hspace{2cm}=\left\{\begin{array}{l}\displaystyle-\frac{(m+1)\hat{T}_0}{\sqrt{4\pi(2m+1)(2m+3)}}
\quad\mathrm{if}\,n=m+1\\
\displaystyle-\frac{m\hat{T}_0}{\sqrt{4\pi(2m+1)(2m-1)}}\quad\mathrm{if}\,n=m-1\\
0\quad\quad\mathrm{otherwise}\end{array}\right.\nonumber\\
&&\{p_{\gamma
n},C[Y_m]\}\approx\hat{T}_0\sqrt{\frac{(2n+1)(2m+1)}{(4\pi)^3}}
\int_{\mathbb{S}^2}\!\!\!^{\dos}\mathrm{e}P_n(\cos\theta)[P_m(\cos\theta)+
\cos\theta P_m^\prime(\cos\theta)]\nonumber\\
&&\hspace{2cm}=\left\{\begin{array}{l}\displaystyle0\quad\quad\mathrm{if}\,\,m=0\,\, \mathrm{or}\,\,m<n\nonumber\\
\displaystyle-\hat{T}_0\frac{n+1}{\sqrt{4\pi}}\quad\mathrm{if}\,\,m=n\\
\displaystyle\star\quad \quad\mathrm{otherwise}\end{array}\right.\nonumber\\
&&\{p_{\gamma
n},C_\theta[Y_m]\}\approx p_{\gamma0}\sqrt{\frac{(2n+1)(2m+1)}{(4\pi)^3}}
\int_{\mathbb{S}^2}\!\!\!^{\dos}\mathrm{e}P_n(\cos\theta)[P_m(\cos\theta)-\sin^2\theta
P_m^\prime(\cos\theta)]\nonumber\\
&&\hspace{2cm}=\left\{\begin{array}{l}\displaystyle0\quad\quad\mathrm{if}\,\,m=0\,\,
\mathrm{or}\,\,m<n-1\nonumber\\
\displaystyle\ast\quad\quad\mathrm{otherwise}\end{array}\right.\nonumber
\end{eqnarray}
where the symbol $\approx$ denotes that we are restricting ourselves
to points in the hypersurface $\Gamma_{G}\subset\Gamma_c$ defined by
the gauge fixing conditions and the constraints. The $\star$ and
$\ast$ symbols denote terms (computable in closed form but with somewhat
complicated expressions) that are not needed in the following
discussion. As before it helps to display the previous result in
table form
\begin{center}
\begin{tabular}{|c|c c|c c|c|}
\hline\hline
&$\hat{T}_1=0$&$p_{\gamma,1}=0$&$\hat{T}_{2}=0$&$p_{\gamma,2}=0$&$\hdots$\\\hline
$C[Y_0]$&$0$&$0$&$0$&$0$&$\hdots$\\
$C_\theta[Y_0]$&$\frac{\hat{T}_0}{2\sqrt{3\pi}}$&$0$&$0$&$0$&$\hdots$\\\hline
$C[Y_1]$&$-\frac{p_{\gamma0}}{2\sqrt{\pi}}$&$\frac{\hat{T}_0}{\sqrt{\pi}}$&$0$&$0$&$\hdots$\\
$C_\theta[Y_1]$&$0$&$\ast$&$\frac{\hat{T}_0}{\sqrt{15\pi}}$&$0$&$\hdots$\\\hline
$C[Y_2]$&$0$&$\star$&$-\frac{p_{\gamma0}}{2\sqrt{\pi}}$&$\frac{3\hat{T}_0}{\sqrt{4\pi}}   $&$\hdots$\\
$C_\theta[Y_2]$&$\frac{\hat{T}_0}{\sqrt{15\pi}}$&$\ast$&$0$&$\ast$&$\hdots$\\\hline
$\vdots$&$\vdots$&$\vdots$&$\vdots$&$\vdots$&$\ddots$\\
\hline\hline
\end{tabular}
\end{center}
One must also check if the polar constraints are gauge fixed by our
conditions (\ref{gauge3S2}). To this end we compute
\begin{eqnarray}
&&\{\hat{T}_n,\hat{T}e^{-\hat{\gamma}/2}\}\approx0\,,\nonumber\\
&&\{p_{\hat{\gamma}n},\hat{T}e^{-\hat{\gamma}/2}\}\approx
\frac{1}{2}\hat{T}e^{-\hat{\gamma}/2}\sqrt{\frac{2n+1}{4\pi}}P_n(\cos\theta)\,.\nonumber
\end{eqnarray}
The last Poisson bracket is different from zero at the poles
$(\theta=0,\pi)$ for all values of $n\in\mathbb{N}$. As we can see
the only constraint that is not gauge-fixed by the conditions
introduced above, as long as $p_{\gamma_0}\neq0$ and
$\hat{T}_0\neq0$, is $C[1]$. This is different
from the situation in the $\mathbb{T}^2$ case where we were left
with two constraints instead of just one.

As we did before we can pullback everything to the phase space
hypersurface defined by the gauge fixing conditions. The induced
2-form becomes
\begin{eqnarray}
\iota^{*}\omega=\mathrm{d}\hat{\gamma}_0\wedge \mathrm{d}
p_{\hat{\gamma}_0}+\mathrm{d}\hat{T}_0\wedge \mathrm{d}
p_{\hat{T}_0}+\sum_{i}\int_{\mathbb{S}^2}{^{\dos}}e\,
\delta\phi_i\wedge \delta p_{\phi_i}\,\label{symplecticS2}
\end{eqnarray}
and the remaining constraint is
\begin{eqnarray}
\label{remCS2}\mathcal{C}&:=&-p_{\hat{\gamma}_0}p_{\hat{T}_0}+\hat{T}_0
\big[4\sqrt{\pi}\big(\log\frac{\hat{T}_0}{\sqrt{4\pi}}-1\big)-\hat{\gamma_0}\big]
\\&+&\frac{1}{2}\sum_{i}\int_{\mathbb{S}^2}\!\!\!^{\dos}\mathrm{e}\left(\frac{\sqrt{4\pi}}{\hat{T}_0}p_{\hat{\phi}_i}^2+
\frac{\hat{T}_0}{\sqrt{4\pi}}\hat{\phi}_i^{\prime2}\sin^2\theta\right)\approx0\,.\nonumber
\end{eqnarray}
The gauge transformations generated by this constraint in the
variables $\hat{T}_0$ and $p_{\hat{\gamma}_0}$ are
\begin{eqnarray}
&&\{\hat{T}_0,\mathcal{C}\}=-p_{\hat{\gamma}_0}\,,\nonumber\\
&&\{p_{\hat{\gamma}_0},\mathcal{C}\}=\hat{T}_0\,,\nonumber
\end{eqnarray}
so if we parameterize the gauge orbits as before with $s\in(0,\pi)$
we find now $\hat{T}_0=p\sin s$ and
$p_{\hat{\gamma}_0}=-p\cos s$, $p\neq0$. In the spirit of the
previous section we introduce now the following canonical
transformation ($(Q,P)$ and $(T,p_{T})$ denote canonically conjugate
pairs)
\begin{eqnarray}
&\displaystyle\hat{T}_0=P\sin T\,,\hspace{1.7cm}&p_{\hat{T}_0}=\frac{p_T}{P}\cos T-Q\sin T\,,\nonumber\\
&\displaystyle\hat{\gamma}_0=-Q\cos T-\frac{p_T}{P}\sin
T\,,&p_{\hat{\gamma}_0}=-P\cos T\,.\nonumber
\end{eqnarray}
In addition, as we did in the $\mathbb{T}^3$ case, it is possible to
write the remaining constraint in a more pleasant form by performing
a further canonical transformation (here, again,
$(\tilde{Q},\tilde{P})$ and $(\varphi_i,p_{\varphi_i})$
are canonical pairs)
\begin{eqnarray}
&\displaystyle\tilde{Q}:=PQ+\frac{1}{2}\sum_{i}\int_{\mathbb{S}^2}{^{\dos}}\mathrm{e}\,p_{\hat{\phi}_i}\hat{\phi}_i\,,
&\displaystyle\tilde{P}:=\log P\,,\nonumber\\
&\displaystyle\varphi_i=(4\pi)^{-1/4}\sqrt{P}\hat{\phi}_i\,,\hspace{1.7cm}&p_{\varphi_i}=(4\pi)^{1/4}\frac{p_{\hat{\phi}_i}}{\sqrt{P}}\,,
\end{eqnarray}
giving
\begin{equation}
\mathcal{C}=p_{T}+4\sqrt{\pi}e^{\tilde{P}}(\log \frac{\sin
T}{\sqrt{4\pi}}+\tilde{P}-1)\sin
T+\frac{1}{2}\sum_{i}\int_{\mathbb{S}^{2}}{^{\dos}}\mathrm{e}\left(\frac{p_{\varphi_{i}}^{2}}{\sin
T}+\varphi_{i}^{\prime2}\sin T\sin^{2}\theta\right)\approx 0\,.
\end{equation}
It is now obvious the interpretation of the system as a
non-autonomous Hamiltonian system
$((0,\pi)\times\Gamma_R,\mathrm{d}t,\omega_R,H_R)$, where
$\Gamma_{R}$ denotes the reduced phase space coordinatized by the
canonical pairs $(\tilde{Q},\tilde{P};\varphi_{i},p_{\varphi_i})$,
endowed with the standard (weakly) simplectic form (\ref{omegaR}).
The dynamics is given by the time dependent Hamiltonian
$H_{R}(t):\Gamma_{R}\rightarrow\mathbb{R}$
\begin{equation}
H_{R}(t)=4\sqrt{\pi}e^{\tilde{P}}(\log \frac{\sin
t}{\sqrt{4\pi}}+\tilde{P}-1)\sin
t+\frac{1}{2}\sum_{i}\int_{\mathbb{S}^{2}}{^{\dos}}\mathrm{e}\left(\frac{p_{\varphi_{i}}^{2}}{\sin
t}+\varphi_{i}^{\prime2}\sin t\sin^{2}\theta\right),\label{4.24}
\end{equation}
with the evolution vector field
\begin{widetext}
\begin{eqnarray}
E_{H_R}&=&\frac{\partial}{\partial
t}+4\sqrt{\pi}e^{\tilde{P}}(\log\frac{\sin
t}{\sqrt{4\pi}}+\tilde{P})\sin
t\frac{\partial}{\partial\tilde{Q}}\nonumber\\
&&+\sum_{i}\int_{\mathbb{S}^{2}}{^{(2)}}\mathrm{e}\left(\frac{p_{\varphi_{i}}}{\sin
t}\frac{\delta}{\delta \varphi_{i}}+(\sin^{2}\theta\varphi_i^{\prime})^{\prime}\sin
t\frac{\delta}{\delta
p_{\varphi_{i}}}\right).
\end{eqnarray}
\end{widetext}
Several comments are in order at this point. First we can see now
that the final description of our system is somewhat simpler that in
the $\mathbb{T}^3$ case because we do not have any remaining
constraints and the fields $\varphi_1$ and $\varphi_2$ are decoupled
(at variance with the previous case). On the other hand we see now
that the dynamics of the global modes, though easy to get in explicit form,
is not as simple as the one found for the torus. Notice also that the
Hamiltonian (\ref{4.24}) is singular whenever $\sin t=0$. This means that if we pick the initial time
$t_0\in(0,\pi)$ in order to write the Cauchy data we meet a past singularity at $t=0$ and a future singularity at $t=\pi$.

\section{$\mathbb{S}^3$ Gowdy models coupled to massless scalars}{\label{S3}}

Let us finally consider the case where the spatial slices
${^{\tres}}\Sigma$ have the topology of a 3-sphere $\mathbb{S}^3$,
described as
$\mathbb{S}^3=\{(z_1,z_2)\in\mathbb{C}^2:|z_1|^2+|z_2|^2=1\}$. Let us
define the following action of $G^{\dos}$ on ${^{\tres}}\Sigma$
\begin{equation}
(g_1,g_2)\cdot(z_1,z_2)=(e^{ix_1},e^{ix_2})\cdot(z_1,z_2)=(e^{ix_1}z_1,e^{ix_2}z_2)\,.
\label{S3action_1}
\end{equation}
The action of the two $U(1)$ subgroup factors is
\begin{eqnarray*}
&&(e^{ix},1)\cdot(z_1,z_2)=(e^{ix}z_1,z_2)\nonumber\\
&&(1,e^{ix})\cdot(z_1,z_2)=(z_1,e^{ix}z_2).\nonumber
\end{eqnarray*}
The corresponding tangent vectors at each point of $\mathbb{S}^3$,
obtained by differentiating the previous expressions with respect to
$x$ at $x=0$, are now
\begin{eqnarray*}
&&(iz_1,0)\nonumber\\
&&(0,iz_2).\nonumber
\end{eqnarray*}
As we can see they vanish at $z_1=0$ and $z_2=0$ (i.e. at the
circumferences $(0,e^{i\xi})$ and $(e^{i\sigma},0)$, $\xi,
\sigma\in\mathbb{R}(\mathrm{mod}\,2\pi)$). This fact poses now the
question of how one can possibly use them to perform a Geroch
reduction that requires us to have at least a non-vanishing Killing
vector field. On some other respects they present no problems, in
particular they are commuting fields. A useful parametrization of
$\mathbb{S}^3$ is $z_1=e^{i\sigma}\sin(\theta/2)$,
$z_2=e^{i\xi}\cos(\theta/2)$ with $\theta\in[0,\pi]$,
$\xi,\sigma\in\mathbb{R}(\mathrm{mod}\,2\pi)$, with the commuting
Killing fields $\sigma^a$ and $\xi^a$ given by
$\sigma^a=(\partial/\partial\sigma)^{a}$ and
$\xi^a=(\partial/\partial\xi)^{a}$. This allows us to view the
three-sphere as a filled torus in which the points on the same
parallel of the surface are identified (so that the surface itself
can be viewed as a circle $\mathbb{S}^1$). This is helpful to
perform the Geroch reduction.

The fact that the Killing vectors that we have chosen vanish
alternatively in two different circles poses a problem as far as the
Geroch reduction is concerned because to perform it one should use a
non-vanishing vector. We will show now that the the fact that
$\xi^{a}$ only vanishes in a one dimensional submanifold will
effectively allow us to use them to carry out this reduction. To this
end we start form an action in four dimensions defined on
${^{\cuatro}}\mathcal{M}$, topologically $\mathbb{R}\times
\mathbb{S}^3$, and remove the circle where the Killing vanishes from
the integration region. As this is a zero-measure set the integral
will not change. Of course one must take now into account the fact
that the fields in the new integration region cannot be completely
arbitrary but should be subject to some restrictions (regularity
conditions) reflecting the fact that they should extend to the full
${^{\cuatro}}\mathcal{M}$ in a smooth way.

Topologically the two dimensional manifold that appears in the action (\ref{act})
is ${^{\dos}}\Sigma=D$, where $D$ denotes an open
disk. The regularity conditions on the disk boundary are such that
the fields (of any tensor type) behave in a ``radial coordinate''
$\theta$ exactly as an axially symmetric field would do in the axis.
Eventually this will allow us to change the disk by a two sphere.

As in the previous cases we are going to use
$(t^{a},\theta^{a},\sigma^{a})$ as coordinate vector fields. We will
write now $\theta^a=f\hat{\theta}^a$,
$t^a=f(Nn^a+N^\theta\hat{\theta}^a)$ with $N>0$ and $f>0$. The
scalars $f$, $N$, and $N^\theta$ are supposed to be smooth fields on
$\mathbb{R}\times D$ subject to some regularity conditions that will
be specified later. Notice that we write now $f$ instead of
$e^\gamma/2$ as in previous cases because we want to allow $f$ to go
to zero at the disk boundary. The same argument that we used for the
two previous cases tells us now that
\begin{eqnarray}
&&N\mathcal{L}_\theta f+f\mathcal{L}_\theta
N+Nf^2n^an^b\nabla_a\hat{\theta}_b=0\,,\label{proy_1_3sph}\\
&&N^\theta\mathcal{L}_\theta f+f\mathcal{L}_\theta
N^\theta-\mathcal{L}_t f+
Nf^2\hat{\theta}^a\hat{\theta}^b\nabla_a n_b=0\,,\label{proy_2_3sph}\\
&&\hat{\theta}^a\sigma^b\nabla_an_b=0\,.\label{proy_3_3sph}
\end{eqnarray}
The form of the 3-metric $g_{ab}$ is basically the same as in the
other cases
\begin{eqnarray*}\label{metric_S3}
g_{ab}=f^2[(N^{\theta 2}-N^{2})(\mathrm{d}t)_a(\mathrm{d}t)_b+
2N^{\theta}(\mathrm{d}t)_{(a}(\mathrm{d}\theta)_{b)}+
(\mathrm{d}\theta)_a(\mathrm{d}\theta)_b]+
\tau^{2}(\mathrm{d}\sigma)_a(\mathrm{d}\sigma)_b\,
\end{eqnarray*}
and the determinant is now given by $|g|=\tau^2N^2f^4$. Again
$(N,N^\theta,\gamma,\phi_{i})$, are constant on the orbits of the
remaining Killing field $\sigma^a$ and hence they only depend on the
coordinates $(t,\theta)$. Using the coordinates system
$(t,\theta,\sigma,\xi)$ we write the original 4-metric
${^{\cuatro}}g_{ab}$ as
\begin{eqnarray}
\label{4metricS3}{^{\cuatro}}g_{ab}&=&\frac{f^2}{\lambda_\xi}[(N^{\theta
2}-N^{2})(\mathrm{d}t)_a(\mathrm{d}t)_b+
2N^{\theta}(\mathrm{d}t)_{(a}(\mathrm{d}\theta)_{b)}+
(\mathrm{d}\theta)_a(\mathrm{d}\theta)_b]\\
&+&
\frac{\tau^{2}}{\lambda_\xi}(\mathrm{d}\sigma)_a(\mathrm{d}\sigma)_b+
\lambda_{\xi}(\mathrm{d}\xi)_a(\mathrm{d}\xi)_b.\nonumber
\end{eqnarray}
We have to find out the regularity conditions satisfied by this
metric. At $\theta=0$ the regularity conditions should be of the
same type as the ones that we have already used in the
$\mathbb{S}^1\times\mathbb{S}^2$ case. Here, however, we also have
to impose regularity conditions when we approach the boundary of
the filled torus that we obtained by removing the circle where the
Killing used to perform the Geroch reduction
vanishes. This can be formally achieved by changing $\theta$ for
$\pi-\theta$. By doing this we find
\begin{eqnarray}
&&\frac{f^2}{\lambda_{\xi}}(N^{\theta 2}-N^2)=A(t,\cos\theta)\,,\label{reg1_S3}\\
&&\frac{f^2}{\lambda_{\xi}}N^\theta=B(t,\cos\theta)\sin\theta\,, \label{reg2_S3}\\
&&\lambda_{\xi}=4\cos^2(\theta/2)[F(t,\cos\theta)-G(t,\cos\theta)\cos^2(\theta/2)]\,,\label{reg3_S3}\\
&&\frac{f^2}{\lambda_{\xi}}=D(t,\cos\theta)+E(t,\cos\theta)\sin^2(\theta/2)=\nonumber\\
&&\hspace{1cm}F(t,\cos\theta)+G(t,\cos\theta)\cos^2(\theta/2)\,,\label{reg4_S3}\\
&&\frac{\tau^2}{\lambda_{\xi}}=4\sin^2(\theta/2)[D(t,\cos\theta)-E(t,\cos\theta)\sin^2(\theta/2)
]\,.\label{reg5_S3}
\end{eqnarray}
Here the functions $A$, $B$, $D$, $E$, $F$, and $G$ are analytic in
their arguments. Notice that they are not independent because they
are constrained to satisfy
$$
D(t,\cos\theta)+E(t,\cos\theta)\sin^2(\theta/2)=
F(t,\cos\theta)+G(t,\cos\theta)\cos^2(\theta/2)\,.
$$
We have used the functions $\sin(\theta/2)$ and $\cos(\theta/2)$
because they alternatively vanish on the circles where the
Killings themselves become zero and have the dependence of a
regular scalar function in terms of the ``radial'' coordinates
$\theta$ or $\pi-\theta$ on the circles where they do not vanish.
The cosine dependence of the other functions is dictated by
regularity at the two circles. This is very important because we
will be able to write down our model in terms of them and, having
$\cos\theta$ as their argument they can be interpreted as functions
on $\mathbb{S}^2$ as in the $\mathbb{S}^1\times\mathbb{S}^2$
case. The conditions that the fields must satisfy (dropping the
$t$-dependence) are
\begin{eqnarray}
&&\lambda_{\xi}=e^{\hat{\phi}_1(\cos\theta)}\cos^2(\theta/2)\,,\label{condfields2_S3}\\
&&\phi_2=\hat{\phi}_2(\cos\theta)\,,\label{condfields1_S3}\\
&&f=\cos(\theta/2)e^{\hat{\gamma}(\cos\theta)/2}\,,\label{condfields3_S3}\\
&&N^\theta=\hat{N}^\theta(\cos\theta)\sin\theta\,,\label{condfields4_S3}\\
&&N=\hat{N}(\cos\theta)\,,\label{condfields5_S3}\\
&&\tau=\hat{T}(\cos\theta)\sin\theta\,,\label{condfields6_S3}\\
&&\hat{T}^2e^{-\hat{\gamma}}=\frac{D(\cos\theta)-E(\cos\theta)\sin^2(\theta/2)}
{D(\cos\theta)+E(\cos\theta)\sin^2(\theta/2)}\,,\label{condfields7_S3}\\
&&e^{2\hat{\phi}_1-\hat{\gamma}}=4\frac{F(\cos\theta)-G(\cos\theta)\cos^2(\theta/2)}
{F(\cos\theta)+G(\cos\theta)\cos^2(\theta/2)}\,,\label{condfields8_S3}
\end{eqnarray}
where we have used $ \sin\theta=2\sin(\theta/2)\cos(\theta/2)$.
Here, as in the $\mathbb{S}^1\times\mathbb{S}^2$ case, we have that
$\hat{\phi}_{i},\hat{\gamma},\hat{N}^{\theta},\hat{N},\hat{T}:[-1,1]\rightarrow\mathbb{R}$
($\hat{N}>0$). They can be written as functions of
$A,\,B,\,D,\,E,\,F,$ and $G$. They must be differentiable in
$(-1,1)$, and their right and left derivatives at $\pm1$ must be
defined (equivalently they must be $\mathcal{C}^\infty$ in $(-1,1)$
with bounded derivative). Conditions
(\ref{condfields7_S3},\ref{condfields8_S3}) imply that
$$\hat{T}(+1)e^{-\hat{\gamma}(+1)/2}=1\,\quad \mathrm{and}\quad
e^{2\hat{\phi}_1(-1)-\hat{\gamma}(-1)}=4\,.$$
These are the polar
constraints for the $\mathbb{S}^3$ topology. This is slightly
different from previous examples because now these conditions
involve different pairs of objets at the two poles of
$\mathbb{S}^2$.

Our starting point is now the action
\begin{eqnarray}\label{act3esf}
\displaystyle {^{\tres}}S(g_{ab},\phi_i)&=&\frac{1}{16\pi
G_{3}}\int_{t_0}^{t_1}\mathrm{d}t\int_{\mathbb{S}^2}{^{\dos}}\mathrm{e}|g|^{1/2}\left(\,^{\dos}\!R+K_{ab}K^{ab}-K^2-
\frac{1}{2}g^{ab}\sum_{i}(\mathrm{d}\phi_{i})_a(\mathrm{d}\phi_{i})_b\right),\nonumber
\end{eqnarray}
with $\phi_1=\log\lambda_{\xi}=\hat{\phi}_1+\log\cos^2(\theta/2)$.
Notice that we have changed the integration region to $\mathbb{S}^2$
because, as we will see, it can be written in terms of the \textit{hat}-fields that are smoothly extended to $\mathbb{S}^2$.

As in the case of the three-handle we choose the fiducial volume
element $^{\dos}\mathrm{e}$ to be compatible with the auxiliary
round metric on the 2-sphere $\mathbb{S}^2$, i.e.
${^{\dos}}\mathrm{e}=\sin\theta\mathrm{d}\theta\wedge\mathrm{d}\sigma$,
with
${^{\dos}}\mathrm{e}_{ab}\theta^{a}\sigma^{b}=Nf^2\tau/|g|^{1/2}=\sin\theta$.
In terms of the fields
$(\hat{N},\hat{N}^{\theta},\hat{\gamma},\hat{T},\hat{\phi}_{i})$ the
action becomes
\begin{eqnarray}
&&\hspace{-.5cm}{^{\tres}}S(\hat{N},\hat{N}^\theta,\hat{\gamma},\hat{T},\hat{\phi}_i)=\nonumber\\
&&\hspace{-.5cm}=\frac{1}{16\pi G_3}\int_{t_0}^{t_1}\!\!\!\!\!
\mathrm{d}t\int_{\mathbb{S}^2}\!\!\!^{\dos}\mathrm{e}\bigg(
\hat{N}[(\hat{\gamma}^\prime\hat{T}^\prime-2\hat{T}^{\prime\prime})\sin^2\theta+
(5\hat{T}^{\prime}-\hat{\gamma}^\prime\hat{T})\cos\theta+\hat{T}^\prime+\frac{3}{2}\hat{T}]\nonumber\\
&&+\frac{1}{\hat{N}}[\hat{N}^\theta
\hat{T}\cos\theta-\dot{\hat{T}}-\hat{N}^\theta
\hat{T}^\prime\sin^2\theta][\dot{\hat{\gamma}}
+(2\hat{N}^{\theta\prime}+\hat{N}^\theta\hat{\gamma}^\prime)\sin^2\theta+(1-3\cos\theta)\hat{N}^\theta]
\label{act_esf}\\
&&+\frac{\hat{T}}{2\hat{N}}\sum_{i}\bigg[\dot{\hat{\phi}}^2_i
+2\hat{N}^\theta\dot{\hat{\phi}}_i\,\hat{\phi}_i^\prime\sin^2\theta+
(\hat{N}^{\theta2}\sin^2\theta-\hat{N}^2)\hat{\phi}_i^{\prime2}\sin^2\theta\bigg]\nonumber\\
&&+\frac{\hat{T}}{2\hat{N}}[2(1-\cos\theta)(\hat{N}^\theta
\dot{\hat{\phi}}_1+(\hat{N}^{\theta2}\sin^2\theta-\hat{N}^2)\hat{\phi}_1^\prime)
+(1-\cos\theta)^2\hat{N}^{\theta2}] \bigg)\,.\nonumber
\end{eqnarray}
It is important to remark at this point that the action is the
integral of a smooth function on the sphere. We arrive at this
result after several non-trivial cancelations of terms that would
diverge at the poles. This reflects the fact that indeed, by
removing the circle where the Killing vector field used in the
Geroch reduction vanishes, we arrive at a consistent description of the model.
It is also worthwhile pointing out that the structure of the
action is very similar to the one found in the
$\mathbb{S}^1\times\mathbb{S}^2$ case but not exactly the same, in
fact we will see later that the differences are important to
guarantee, for example, the stability of the polar constraints in
this case.

The Hamiltonian of the system can be readily obtained. As in
previous cases it can be written as a sum of constraints
$H=C[\hat{N}]+C_\theta[\hat{N}^\theta]$ with
\begin{eqnarray}
C[\hat{N}]&=& \int_{\mathbb{S}^2}{^{\dos}}\mathrm{e}\hat{N}C
\\
&
&\hspace{-1.2cm}=\int_{\mathbb{S}^2}\!\!\!^{\dos}\mathrm{e}\hat{N}\left(
-16\pi G_3 p_{\hat{\gamma}}p_{\hat{T}}+\frac{1}{16\pi
G_3}\big[(2\hat{T}^{\prime\prime}-\hat{\gamma}^\prime\hat{T}^\prime)\sin^2\theta
+(\hat{\gamma}^\prime\hat{T}-5\hat{T}^\prime)\cos\theta-\frac{3}{2}\hat{T}-\hat{T}^\prime\big]\right.\nonumber\\
&&+\frac{1}{2}\sum_{i}\bigg(\frac{16\pi
G_3p_{\hat{\phi}_i}^2}{\hat{T}} +\frac{\hat{T}}{16\pi
G_3}\hat{\phi}_i^{\prime2}\sin^2\theta\bigg)
+(1-\cos\theta)\frac{\hat{T}}{16\pi G_3}\hat{\phi}_1^\prime\bigg)\nonumber\\
\label{esc_const_S3} C_\theta[\hat{N}^\theta]&=&
\int_{\mathbb{S}^2}{^{\dos}}\mathrm{e}\hat{N}^{\theta}C_{\theta}
\\
&&\hspace{-1.2cm}
=\int_{\mathbb{S}^2}\!\!\!^{\dos}\mathrm{e}\hat{N}^\theta
\bigg([2p^\prime_{\hat{\gamma}}-\hat{\gamma}^\prime
p_{\hat{\gamma}}- \hat{T}^\prime
p_{\hat{T}}-\sum_{i}\hat{\phi}_i^\prime
p_{\hat{\phi}_i}]\sin^2\theta +[\hat{T}
p_{\hat{T}}-p_{\hat{\gamma}}+p_{\hat{\phi_1}}]\cos\theta-p_{\hat{\gamma}}-p_{\hat{\phi}_1}\bigg).
\nonumber
\end{eqnarray}
The two previous expressions, together with the conditions at the
poles $\hat{T}(+1)e^{-\hat{\gamma}(+1)/2}=1$ and
$e^{2\hat{\phi}_1(-1)-\hat{\gamma}(-1)}=4$, define the constraints
of the system. As before, the polar constraints are necessary
conditions to guarantee the differentiability of the (weighted)
constraints $C[N_g]$ and $C_\theta[N_g^\theta]$. The gauge
transformations defined by $C[\hat{N}_g]$ and
$C_\theta[\hat{N}_g^\theta]$ are\footnote{Again we take $16\pi
G_3=1$.}
\begin{eqnarray*}
&&\{\hat{\gamma},C[\hat{N}_g]\}=-\hat{N}_g p_{\hat{T}}\,,
\nonumber\\
&&\{\hat{T},C[\hat{N}_g]\}=-\hat{N}_g p_{\hat{\gamma}}\,,\nonumber\\
&&\{\hat{\phi_i},C[\hat{N}_g]\}=\hat{N}_g\frac{p_{\hat{\phi}_i}}{\hat{T}}\,,\nonumber\\
&&\{p_{\hat{\gamma}},C[\hat{N}_g]\}=\hat{N}_{g}^{\prime}(\hat{T}\cos\theta-\hat{T}^\prime\sin^2\theta)
+\hat{N}_g(3\hat{T}^\prime\cos\theta+\hat{T}-\hat{T}^{\prime\prime}\sin^2\theta),\nonumber\\
&&\{p_{\hat{T}},C[\hat{N}_g]\}=
\hat{N}_g\big[\frac{1}{2}-\hat{\phi}_1^\prime+(\hat{\gamma}^\prime+\hat{\phi}_1^\prime)\cos\theta
-\hat{\gamma}^{\prime\prime}\sin^2\theta\big]+
\hat{N}_g^\prime(3\cos\theta-1-\hat{\gamma}^\prime\sin^2\theta)\nonumber\\
&&\hspace{2.5cm}-
2\hat{N}_g^{\prime\prime}\sin^2\theta+\frac{\hat{N}_g}{2}\sum_{i}\bigg(\frac{p_{\hat{\phi}_i}^2}{\hat{T}^2}
-\sin^2\theta\hat{\phi}_i^{\prime2}\bigg)\,,\nonumber\\
&&\{p_{\hat{\phi}_1},C[\hat{N}_g]\}=[\hat{N}_g\hat{T}(\hat{\phi}^\prime_2\sin^2\theta+1-\cos\theta)]^\prime\,,\nonumber\\
&&\{p_{\hat{\phi}_2},C[\hat{N}_g]\}=
(\hat{N}_g\hat{T}\hat{\phi}^\prime_2\sin^2\theta)^\prime\,,\nonumber
\end{eqnarray*}
and
\begin{eqnarray*}
&&\{\hat{\gamma},C_\theta[\hat{N}^\theta_g]\}=-2\hat{N}_g^{\theta\prime}\sin^2\theta+\hat{N}_g^\theta(3\cos\theta-\hat{\gamma}^\prime\sin^2\theta-1)\,,\nonumber\\
&&\{\hat{T},C_\theta[\hat{N}^\theta_g]\}=\hat{N}_g^\theta(\hat{T}\cos\theta-\hat{T}^\prime\sin^2\theta)\,,\nonumber\\
&&\{\hat{\phi}_1,C_\theta[\hat{N}^\theta_g]\}=\hat{N}_g^\theta(\cos\theta-1-\hat{\phi}_2^\prime\sin^2\theta)\,,\nonumber\\
&&\{\hat{\phi}_2,C_\theta[\hat{N}^\theta_g]\}=-\hat{N}_g^\theta\hat{\phi}_2^\prime\sin^2\theta\,,\nonumber\\
&&\{p_{\hat{\gamma}},C_\theta[\hat{N}^\theta_g]\}=-(\hat{N}_g^\theta p_{\hat{\gamma}}\sin^2\theta)^\prime\,,\nonumber\\
&&\{p_{\hat{T}},C_\theta[\hat{N}^\theta_g]\}=\hat{N}_g^\theta(p_{\hat{T}}\cos\theta-p_{\hat{T}}^\prime\sin^2\theta)
-\hat{N}_g^{\theta\prime}p_{\hat{T}}\sin^2\theta\,,
\nonumber\\
&&\{p_{\hat{\phi}_i},C_\theta[\hat{N}^\theta_g]\}=-(\hat{N}_g^\theta p_{\hat{\phi_i}}\sin^2\theta)^\prime\,.\nonumber\\
\end{eqnarray*}
The Poisson brackets of these constraints give exactly the same
result that we obtained for the $\mathbb{S}^1\times\mathbb{S}^2$
topology and, hence, define a fist class constrained surface
$\Gamma_c\subset\Gamma$. Here $(\Gamma,\omega)$ denotes the
canonical phase space of the system, coordinatized by the
canonical pairs
$(\hat{\gamma},p_{\hat{\gamma}};\hat{T},p_{\hat{T}};\hat{\phi}_{i},p_{\hat{\phi}_i})$,
and endowed with the standard (weakly) symplectic form
(\ref{omegaS2XS1}). We must check now the stability of the
polar constraints. We do this by computing
\begin{eqnarray}
&&\{\hat{T}e^{-\hat{\gamma}/2},C[\hat{N}_g]\}=\frac{1}{2}\hat{N}_ge^{-\hat{\gamma}/2}(\hat{T}p_{\hat{T}}-2p_{\hat{\gamma}})\,,\label{pb1}\\
&&\{\hat{T}e^{-\hat{\gamma}/2},C_\theta[\hat{N}_g^\theta]\}=e^{-\hat{\gamma}/2}
\bigg(\frac{1}{2}\hat{N}_g^\theta\hat{T}(1-\cos\theta)+(\hat{N}_g^{\theta\prime}\hat{T}
-\hat{N}_g^\theta\hat{T}^\prime+\frac{1}{2}\hat{N}_g^\theta\hat{T}\hat{\gamma}^\prime)\sin^2\theta\bigg),\quad\quad\label{pb2}\\
&&\{e^{2\hat{\phi}_1-\hat{\gamma}},C[\hat{N}_g]\}=\frac{\hat{N}^g}{\hat{T}}
e^{2\hat{\phi}_1-\hat{\gamma}}(2p_{\hat{\phi}_1}+\hat{T}p_{\hat{T}})\,,\label{pb3}\\
&&\{e^{2\hat{\phi}_1-\hat{\gamma}},C_\theta[\hat{N}_g^\theta]\}=e^{2\hat{\phi}_1-\hat{\gamma}}
\Big(-\hat{N}_g^\theta(1+\cos\theta)+
(2\hat{N}_g^{\theta\prime}-2\hat{N}_g^\theta\hat{\phi}_2^\prime
+\hat{N}_g^\theta\hat{\gamma}^\prime)\sin^2\theta\Big)\,.\label{pb4}
\end{eqnarray}
The constraint (\ref{esc_const_S3}) at the poles $\theta=0,\pi$
gives respectively,
$\hat{T}(+1)p_{\hat{T}}(+1)-2p_{\hat{\gamma}}(+1)=0$, and
$\hat{T}(-1)p_{\hat{T}}(-1)+2p_{\hat{\phi}_1}(-1)=0$. These
guarantee that the Poisson bracket (\ref{pb1}), vanishes at
$\theta=0$ and (\ref{pb3}) vanishes at $\theta=\pi$. The vanishing
of (\ref{pb2}) at $\theta=0$ is due to the presence of the factors
$1-\cos\theta$ and $\sin^2\theta$ and, finally, (\ref{pb4}) is zero
at $\theta=\pi$ due to the factors $1+\cos\theta$ and
$\sin^2\theta$. As in the $\mathbb{S}^1\times\mathbb{S}^{2}$ we
conclude that there are no secondary constraints coming from the

stability of these polar constraints.

The deparameterization in this case follows closely the one for
$\mathbb{S}^1\times\mathbb{S}^2$. The same gauge fixing conditions
work in our case now. The only new element now is checking if the
polar constraints are gauge fixed or not and this only requires
the computation of
\begin{eqnarray*}
&&\{p_{\hat{\gamma}n},e^{2\hat{\phi}_1-\hat{\gamma}}\}=
e^{2\hat{\phi}_1-\hat{\gamma}}\sqrt{\frac{2n+1}{4\pi}}P_n(\cos\theta)
\end{eqnarray*}
which is different from zero at the poles. As we see the situation
now is completely analogous to the previous case. The pull-back of
the symplectic form to the phase space hypersurface defined by the
gauge fixing conditions is given again by (\ref{symplecticS2}). We
are left only with the constraint
\begin{eqnarray}
\label{constS3}\mathcal{C} &:=& -p_{\hat{\gamma}_{0}}p_{\hat{T}_{0}}+\hat{T}_{0}\left(\sqrt{4\pi}
\log\frac{\hat{T}_{0}}{\sqrt{4\pi}}-\hat{\gamma}_{0}-\sqrt{\pi}(2\log2+3)
+\hat{\phi}_{1_{0}}\right)\\
&+&\frac{1}{2}\sum_{i}\int_{\mathbb{S}^{2}}
{^{\scriptstyle{\dos}}}\mathrm{e}\left(\frac{\sqrt{4\pi}p_{\hat{\phi}_{i}}^{2}}{\hat{T}_{0}}
+\frac{\hat{T}_{0}}{\sqrt{4\pi}}\hat{\phi}_i^{\prime2}\right)\approx0\,.\nonumber
\end{eqnarray}
The gauge transformations generated by this constraint on the
variables $\hat{T}_0$ and $p_{\hat{\gamma}0}$ are the same as for
the three-handle and, hence, we can use the canonical
transformations introduced at the end of the previous section to
rewrite (\ref{constS3}) as
\begin{eqnarray}
\label{finconstS3}p_T&+&(4\pi)^{1/4}e^{\tilde{P}/2}\hat{\varphi}_{1_0}\sin
T+2\sqrt{\pi}e^{\tilde{P}}(\log \frac{\sin
T}{\sqrt{4\pi}}+\tilde{P}-\log2-\frac{3}{2})\sin
T\\
&+&\frac{1}{2}\sum_{i}\int_{\mathbb{S}^2}\!\!\!^{\dos}\mathrm{e}
\bigg(\frac{p_{\hat{\varphi}_i}^2}{\sin
T}+\hat{\varphi}_i^{\prime2}\sin
T\sin^2\theta\bigg)\approx0.\nonumber
\end{eqnarray}
The description of the system by a time-dependent Hamiltonian is now
straightforward. It is interesting at this point to compare the
dynamics of this model and the $\mathbb{S}^1\times\mathbb{S}^2$ one.
First of all we see that the global mode have a different behavior
now, in particular couples to $\varphi_{1_0}$ through the
term $e^{\tilde{P}/2}\hat{\varphi}_{1_0}\sin T$ in
(\ref{finconstS3}). As we see the gravitational and matter modes
cease to play a symmetric role in this particular description, at
variance with the other topologies. However, by writing the
regularity condition (\ref{condfields2_S3}) with an extra $\hat{T}$
(as will be justified in the next section) it is possible to restore
the symmetry between the gravitational and matter scalars in a
straightforward way.

As in the previous cases, it is possible to interpret the system as
a non-autonomous Hamiltonian system
$((0,\pi)\times\Gamma_R,\mathrm{d}t,\omega_R,H_R)$, where
$\Gamma_{R}$ denotes the reduced phase space coordinatized by the
canonical pairs $(\tilde{Q},\tilde{P};\varphi_{i},p_{\varphi_i})$,
endowed with the standard (weakly) simplectic form (\ref{omegaR}).
The dynamics is given by the time dependent Hamiltonian
$H_{R}(t):\Gamma_{R}\rightarrow\mathbb{R}$
\begin{eqnarray*}
H_{R}(t)&=&(4\pi)^{1/4}e^{\tilde{P}/2}\varphi_{1_0}\sin
t+2\sqrt{\pi}e^{\tilde{P}}(\log \frac{\sin
t}{\sqrt{4\pi}}+\tilde{P}-\log2-\frac{3}{2})\sin
t\\
&+&\frac{1}{2}\sum_{i}\int_{\mathbb{S}^2} {^{\dos}}\mathrm{e}\,
\bigg(\frac{p_{\varphi_i}^2}{\sin t}+\varphi_i^{\prime2}\sin
t\sin^2\theta\bigg),
\end{eqnarray*}
with the evolution vector field given by
\begin{eqnarray*}
E_{H_R}&=&\frac{\partial}{\partial
t}+\left[(4\pi)^{1/4}e^{\tilde{P}/2}\varphi_{1_0}\sin
t+2\sqrt{\pi}e^{\tilde{P}}\sin t\left(\log\frac{\sin
t}{\sqrt{4\pi}}+\tilde{P}-\log2-\frac{1}{2}\right)\right]
\frac{\partial}{\partial\tilde{Q}}\\
&-&(4\pi)^{1/4}e^{\tilde{P}/2} \sin t\frac{\partial}{\partial
p_{\varphi_{1_0}}}+\sum_{i}\int_{\mathbb{S}^{2}}{^{(2)}}\mathrm{e}\left(\frac{p_{\varphi_{i}}}{\sin
t}\frac{\delta}{\delta
\varphi_{i}}+(\sin^{2}\theta\varphi_i^{\prime})^{\prime}\sin
t\frac{\delta}{\delta p_{\varphi_{i}}}\right).
\end{eqnarray*}
The singularities in this case show up in the same way as for the $\mathbb{S}^1\times\mathbb{S}^2$ topology.

\section{Gowdy models as scalar field theories in  2+1 curved
background}{\label{back}}

The purpose of this section is to reinterpret the reduced models
presented in the previous sections as certain simple massless scalar
field theories in conformally stationary backgrounds. We will show
how the metrics obtained after the specific gauge fixing and
deparameterization used in the previous sections can be employed to
reinterpret the meaning (and solution) of the field equations for
each topology. This will allow us to use well-known techniques of
quantum field theory in curved backgrounds to quantize these systems
\cite{BarberoG.:2007}.

Let us start by giving a simple way to solve equations (\ref{ecs})
\begin{eqnarray}
&&R_{ab}=\frac{1}{2}\sum_i
(\mathrm{d}\phi_i)_a(\mathrm{d}\phi_i)_b\,,\label{e1}\\
&&g^{ab}\nabla_a\nabla_b\phi_i=0\,,\label{e2}\\
&&\mathcal{L}_\sigma\phi_i=0.\label{e3}
\end{eqnarray}
If a specific solution
$(\mathring{g}_{ab},\mathring{\phi}_1,\mathring{\phi}_2)$ is known
it is possible to decouple (\ref{e1}) and (\ref{e2},\ref{e3})
because, when (\ref{e3}) is satisfied, we have the equivalence
\begin{eqnarray*}
g^{ab}\nabla_a\nabla_b\phi_i=0 \Leftrightarrow \mathring{g}^{ab}
\mathring{\nabla}_a\mathring{\nabla}_b\phi_i=0.
\end{eqnarray*}
The idea is then to solve the last equation in the background
$\mathring{g}_{ab}$ and then equation (\ref{e1}) just gives
integrability conditions allowing us to recover $g_{ab}$. We will
discuss next the specific form of $\mathring{g}_{ab}$ for each of
the spatial topologies considered in the paper.

\begin{itemize}

\item[$\bullet$] \textbf{Background metric for $\mathbb{T}^3$}

In this case the form of the metric $g_{ab}$ found after the
deparameterization is
$$
g_{ab}=e^\gamma\Big(-(\mathrm{d}t)_a(\mathrm{d}t)_b+(\mathrm{d}\theta)_a(\mathrm{d}\theta)_b\Big)
+\frac{P^2t^2}{2\pi}(\mathrm{d}\sigma)_a(\mathrm{d}\sigma)_b
$$
defined on $(0,\infty)\times \mathbb{T}^2$. A possible (non
unique) choice for
$(\mathring{g}_{ab},\mathring{\phi}_1,\mathring{\phi}_2)$ is
\begin{eqnarray*}
\mathring{g}_{ab}&=&t^2\Big(-(\mathrm{d}t)_a(\mathrm{d}t)_a+(\mathrm{d}\theta)_a(\mathrm{d}\theta)_b
+(\mathrm{d}\sigma)_a(\mathrm{d}\sigma)_b\Big)\\
\mathring{\phi}_1&=&\log t\\
\mathring{\phi}_2&=&0
\end{eqnarray*}
where it is important to notice that even though $\mathring{g}_{ab}$
is not stationary it is conformal to a (flat) stationary metric on
$(0,\infty)\times\mathbb{T}^2$.

\item[$\bullet$] \textbf{Background metric for $\mathbb{S}^1\times\mathbb{S}^2$ }

After deparameterization we get now
$$
g_{ab}=e^\gamma\Big(-(\mathrm{d}t)_a(\mathrm{d}t)_b+(\mathrm{d}\theta)_a(\mathrm{d}\theta)_b\Big)+
\frac{P^2}{4\pi}\sin^2 t
\sin^2\theta(\mathrm{d}\sigma)_a(\mathrm{d}\sigma)_b
$$
defined on $(0,\pi)\times  \mathbb{S}^2$. In this case a
convenient choice for
$(\mathring{g}_{ab},\mathring{\phi}_1,\mathring{\phi}_2)$ is
\begin{eqnarray*}
\mathring{g}_{ab}&=&\sin^2t\Big(-(\mathrm{d}t)_a(\mathrm{d}t)_a+(\mathrm{d}\theta)_a(\mathrm{d}\theta)_b+\sin^2\theta(\mathrm{d}\sigma)_a(\mathrm{d}\sigma)_b\Big)\\
\mathring{\phi}_1&=&\log\sin(t/2)-\log\cos(t/2)\\
\mathring{\phi}_2&=&0.
\end{eqnarray*}
Again this metric is not stationary but it is equal to a time
dependent conformal factor times the Einstein static metric on
$(0,\pi)\times\mathbb{S}^2$.

\item[$\bullet$] \textbf{Background metric for $\mathbb{S}^3$}

Finally we have now
$$
 g_{ab}=\cos^2(\theta/2)e^\gamma\bigg(-(\mathrm{d}t)_a(\mathrm{d}t)_b
+(\mathrm{d}\theta)_a(\mathrm{d}\theta)_b\bigg)+\frac{P^2}{4\pi}\sin^2t\sin^2\theta
(\mathrm{d}\sigma)_a(\mathrm{d}\sigma)_b
$$
defined on $(0,\pi)\times D$ where $D$ denotes the open disk
introduced in the previous section. In this case, a possible choice
of $(\mathring{g}_{ab},\mathring{\phi}_1,\mathring{\phi}_2)$ is
\begin{eqnarray*}
\mathring
g_{ab}&=&\cos^2(\theta/2)e^{\mathring{\gamma}}\bigg(-(\mathrm{d}t)_a(\mathrm{d}t)_b
+(\mathrm{d}\theta)_a(\mathrm{d}\theta)_b\bigg)+\sin^2t\sin^2\theta (\mathrm{d}\sigma)_a(\mathrm{d}\sigma)_b\\
\mathring{\phi}_1&=&\cos\theta \cos t\log(\tan(t/2))+\cos\theta+\log(\cos^2(\theta/2))+\log(2\sin t)\\
\mathring{\phi}_2&=&0
\end{eqnarray*}
where
\begin{eqnarray*}
\mathring{\gamma}&=& \frac{\sin^2\theta}{4}\Big(\sin^2 t\log^2(\tan
t/2)-2\cos t\log(\tan t/2)-1\Big)+\log\left(\sin^2 t\right)\\
&-&\cos t\log(\tan(t/2))+\cos\theta \cos
t\log(\tan(t/2))+\cos\theta -1.
\end{eqnarray*}
It is important to realize that the concrete functional form of
$\mathring{\gamma}$ is irrelevant because, whenever
$\mathcal{L}_\sigma \phi_i=0$, we have the following equivalence in
$(0,\pi)\times(\mathbb{S}^2-\{\theta=\pi\})$
$$
\mathring g^{ab}\mathring \nabla_a\mathring \nabla_b
\phi_i=0\Leftrightarrow \breve{g}^{ab} \breve{\nabla}_a
\breve{\nabla}_b \phi_i=0
$$
with
$$
\breve{g}_{ab}=\sin^2t\Big(-(\mathrm{d}t)_a(\mathrm{d}t)_b
+(\mathrm{d}\theta)_a(\mathrm{d}\theta)_b+\sin^2\theta(\mathrm{d}\sigma)_a(\mathrm{d}\sigma)_b\Big)\,.
$$
Notice that the metric $ \breve{g}_{ab}$ is the one that we found
for $(0,\pi)\times \mathbb{S}^2$ restricted to the manifold
$(0,\pi)\times D$ obtained by removing a point from the sphere.

It is important to point out that $\phi_1$ cannot be extended to
the boundary of the disk, parameterized as $\theta=\pi$,  because
(\ref{condfields2_S3}) forces $\phi_1$ to behave as
$\log(\cos^2(\theta/2))$ for $\theta\rightarrow\pi$. However if we
split $\phi_1$ as
$\phi_1=\phi_1^{\mathrm{sing}}+\phi_1^{\mathrm{reg}}$ with $
\phi_1^{\mathrm{sing}}=\log(\cos^2(\theta/2))+\log(2\sin t)$,
satisfying
$$
\breve{g}^{ab} \breve{\nabla}_a \breve{\nabla}_b
\phi^{\mathrm{sing}}_1=0\,,
$$
we guarantee that the degrees of freedom  contained in
$\phi^{\mathrm{reg}}_1$ still satisfy $\breve{g}^{ab}
\breve{\nabla}_a \breve{\nabla}_b \phi_1^{\mathrm{reg}}=0$ (just the
same equation as the matter field $\phi_2$) and can be extended to
$(0,\pi)\times \mathbb{S}^2$. Notice that the role of the scalar
fields $\phi_1^{\mathrm{reg}}$ and $\phi_2$, both well behaved on
$(0,\pi)\times\mathbb{S}^2$, is symmetric just as in the description
of the previous topologies.

\end{itemize}

\bigskip

 It is important to notice that the scalar field dynamics
generated by the time dependent Hamiltonians that we have obtained
in the previous sections corresponds exactly to the one defined by
the Klein-Gordon equations on the backgrounds given by
$\mathring{g}_{ab}$.

\bigskip

To end this section we want to point out that there are certain
obstructions to the unitary implementation of quantum dynamics for
these systems. Specifically, it can be shown that it is impossible to
find a Fock space representation in which time evolution is
unitarily implementable \cite{Corichi:2002vy,Torre:2002xt}. The
solution to this problem for the torus case relies on certain field
redefinitions involving functions of time
\cite{Corichi:2006xi,Corichi:2006zv}. These can be understood in the
present scheme as coming from the time dependent conformal factors
appearing in $\mathring{g}_{ab}$ (or $\breve{g}_{ab}$). As we will
show in a forthcoming paper, the solution to the unitarity problem
for the topologies considered here relies on field redefinitions
involving precisely the conformal factors shown above. Indeed,
by performing a redefinition of the scalar fields at the Lagrangian
level, such that the conformal factor relating both metrics is traded
by a time-dependent potential term, we expect to find unitary dynamics
if this potential is well behaved. In these cases the background metric
corresponds to a simple, fixed stationary background.

\section{Conclusions and comments}{\label{conclusions}}

We have studied in this paper the Hamiltonian formalism for the
compact, linearly polarized Gowdy models coupled to massless scalar
fields. The purpose of the analysis is to have a Hamiltonian
formulation of the models that can be a starting point for their
quantization either \textit{\`a la Dirac} or by gauge fixing and
deparameterization. The results for the $\mathbb{T}^3$ topology
reproduce the known ones for the gravitational sector and show that
the interaction of the matter fields occur though the constraints
left over by the deparameterization of the system. In the other two
cases the coupling of matter and gravity degrees of freedom can only
be seen when the four metric is reconstructed.

The description of the $\mathbb{S}^1\times\mathbb{S}^2$  and
$\mathbb{S}^3$ models requires a careful discussion of the
regularity conditions that the metric must satisfy in the symmetry
axis left over after the Geroch reduction performed to
describe the systems in 2+1 dimensions. These regularity
conditions are responsible for the appearance of the so called
\textit{polar constraints}. These can be shown to be first class
and are necessary conditions to guarantee the differentiability of
the other constraints present in the models. Of course they must
be taken into account in a Dirac quantization of the Gowdy models
corresponding to these topologies.

An interesting feature of both the
$\mathbb{S}^1\times\mathbb{S}^2$ and $\mathbb{S}^3$ cases is the
fact that after the deparameterization introduced in the paper
(which is a straightforward generalization of the ones used in
the literature for the familiar $\mathbb{T}^3$ case) there are no
constraints left so that the system can be completely described by
the time dependent Hamiltonians that we have found. This is in
contrast with the situation for the 3-torus where in addition to
the dynamics generated by the time dependent Hamiltonian there is
an additional constraint in the system that must be appropriately
taken into account.

A somewhat surprising fact is the possibility to describe both the
$\mathbb{S}^1\times\mathbb{S}^2$  and $\mathbb{S}^3$ models by
using smooth fields on the sphere $\mathbb{S}^2$. An interesting
possibility that may teach us something in the case of
$\mathbb{S}^3$ is to use a Hopf fibration to perform the Geroch
reduction to get a 2+1 dimensional description. This may be the
subject of future work.

The dynamics of the global modes for the different spatial
topologies is easy to obtain but there are significant differences
depending on the topologies. Whereas in the $\mathbb{T}^3$ case the
value of $\tilde{Q}$ and $\tilde{P}$ are just constants in the other
cases $\tilde{P}$ is constant but $\tilde{Q}$ is a function of time.

We have been able to understand in very simple terms the
appearance of both initial and final singularities in the
spacetime metrics that solve the Einstein-Klein Gordon equations
for these models from the point of view of the phase space
description of the dynamics, in particular after the
deparameterization process that we have followed. As we have seen
there are natural variables with very simple gauge transformations
(``time dynamics'') that suggest canonical transformations that
lead to the time dependent Hamiltonians describing the dynamics
and explicitly show how the singularities appear. In the
$\mathbb{S}^1\times\mathbb{S}^2$  and $\mathbb{S}^3$ topologies
the function $\sin t$ in a denominator of the time-dependent
Hamiltonian shows that both final and initial singularities are
present whereas the $t$ denominator in the Hamiltonian for the
3-torus shows that only an initial (or final) singularity appears
in this case.

\begin{acknowledgments}
The authors want to thank I. Garay and J. M. Mart\'{\i}n Garc\'{\i}a
for discussions. Daniel G\'omez Vergel acknowledges the support of
the Spanish Research Council (CSIC) through a I3P research
assistantship. This work is also supported by the Spanish MEC under
the research grant FIS2005-05736-C03-02.

\end{acknowledgments}

\end{document}